\documentclass[aps, preprint, superscriptaddress]{revtex4-2}
\usepackage{graphicx}
\usepackage{tabularx}
\usepackage{makecell, multirow}
\usepackage{color, bm, soul, xcolor}
\usepackage{amsmath, amssymb}
\usepackage{gensymb}
\usepackage{geometry}
\usepackage{braket}
\usepackage{lineno}
\usepackage[colorlinks=true, linkcolor=black, citecolor=black, urlcolor=blue]{hyperref}
\newcommand{\Vo}{{$V_{\rm O}^0$}}
\newcommand{\Vp}{{$V_{\rm O}^{2+}$}}
\newcommand{\Op}{{O$^{\rm p}$}}
\newcommand{\Onp}{{O$^{\rm np}$}}

\begin{document}


\title{Progress in Computational Understanding of Ferroelectric Mechanisms in HfO$_2$}

\author{Tianyuan Zhu}
\affiliation{Key Laboratory for Quantum Materials of Zhejiang Province, Department of Physics, School of Science, Westlake University, Hangzhou, Zhejiang 310024, China}
\affiliation{Institute of Natural Sciences, Westlake Institute for Advanced Study, Hangzhou, Zhejiang 310024, China}

\author{Liyang Ma}
\affiliation{Key Laboratory for Quantum Materials of Zhejiang Province, Department of Physics, School of Science, Westlake University, Hangzhou, Zhejiang 310024, China}

\author{Shiqing Deng}
\affiliation{Beijing Advanced Innovation Center for Materials Genome Engineering, University of Science and Technology Beijing, Beijing 100083, China}

\author{Shi Liu}
\email{liushi@westlake.edu.cn}
\affiliation{Key Laboratory for Quantum Materials of Zhejiang Province, Department of Physics, School of Science, Westlake University, Hangzhou, Zhejiang 310024, China}
\affiliation{Institute of Natural Sciences, Westlake Institute for Advanced Study, Hangzhou, Zhejiang 310024, China}

\begin{abstract}{
Since the first report of ferroelectricity in nanoscale HfO$_2$-based thin films in 2011, this silicon-compatible binary oxide has quickly garnered intense interest in academia and industry, and continues to do so. Despite its deceivingly simple chemical composition, the ferroelectric physics supported by HfO$_2$ is remarkably complex, arguably rivaling that of perovskite ferroelectrics. Computational investigations, especially those utilizing first-principles density functional theory (DFT), have significantly advanced our understanding of the nature of ferroelectricity in these thin films. In this review, we provide an in-depth discussion of the computational efforts to understand ferroelectric hafnia, comparing various metastable polar phases and examining the critical factors necessary for their stabilization. The intricate nature of HfO$_2$ is intimately related to the complex interplay among diverse structural polymorphs, dopants and their charge-compensating oxygen vacancies, and unconventional switching mechanisms of domains and domain walls, which can sometimes yield conflicting theoretical predictions and theoretical-experimental discrepancies. We also discuss opportunities enabled by machine-learning-assisted molecular dynamics and phase-field simulations to go beyond DFT modeling, probing the dynamical properties of ferroelectric HfO$_2$ and tackling pressing issues such as high coercive fields.
}
\end{abstract}

\maketitle

\clearpage

\section{INTRODUCTION}

The discovery of ferroelectricity in silicon-doped hafnium oxide (HfO$_2$) thin films~\cite{Boscke11p102903} quickly positioned HfO$_2$ as a prime material for incorporating ferroelectric functionalities into integrated circuits. Its robust switchable electrical polarization at the nanoscale~\cite{Lee20p1343, Cheema20p478}, simple chemical composition, and excellent compatibility with the modern complementary metal oxide semiconductor (CMOS) technology have led to its implementation in various memory devices~\cite{Luo20p1391,Kim21peabe1341}, sparking renewed interest in developing nonvolatile, energy-efficient, and high-speed ferroelectric memory. Moreover, this discovery has fundamentally enhanced our understanding of ferroelectric physics, which were traditionally focused on perovskite oxides. There are several notable conceptual breakthroughs stemming from the discovery of ferroelectricity in hafnia. First, the prototypical perovskite ferroelectric, BaTiO$_3$, contains the $3d$ element titanium, where the Ti $3d$--O $2p$ hybridization is crucial for stabilizing spontaneous displacements of $d^0$ Ti$^{4+}$ cations~\cite{Cohen92p136}. In contrast, Hf is a $5d$ element and the Hf--O bonds are largely ionic in nature, contributing to the large band gap of HfO$_2$. Second, it has been found that the polar phases of HfO$_2$ are energetically higher than the nonpolar ground-state phase in bulk, highlighting the importance of exploring metastable ferroelectric materials. Last but not least, unlike primarily cation-driven ferroelectricity in perovskite oxides, the inversion symmetry breaking in HfO$_2$'s polar phases is mainly due to local displacements of oxygen anions. This distinction is crucial for understanding why hafnia-based ferroelectrics are more susceptible to oxygen vacancies compared to their perovskite counterparts. It also opens the possibility of bypassing the well-known $d^0$ requirement~\cite{Hill00p6694} of ferroelectric perovskites, potentially leading to the design of new multiferroic materials~\cite{Yu23p142902, Yu23p8127} that combine ferroelectricity with ferromagnetism or even superconductivity~\cite{Duan23pL241114}. 

Several review articles have already summarized different aspects of ferroelectric hafnia, including the properties and origin of ferroelectricity in HfO$_2$-based materials~\cite{Schroeder22p653}, their applications~\cite{Mikolajick21p100901}, the effects of oxygen defects and dopants~\cite{Materano21p2650}, and issues related to device reliability and performance~\cite{Jiang20p2000728, Ihlefeld22p240502}. This review is dedicated to the computational efforts undertaken to decipher the ferroelectric mechanisms in HfO$_2$. Despite its chemical simplicity, HfO$_2$ hosts a plethora of complexities, some of which uniquely differ from perovskite ferroelectrics. This review begins with an examination of hafnia polymorphs in the section `Diverse structural polymorphs', where we compare various polar phases that have been proposed to explain the ferroelectricity observed in thin films produced by different techniques. The metastable nature of the polar phases of HfO$_2$ has led to debates over the exact symmetry of the ferroelectric phase. By carefully analyzing numerous studies in this field, we aim to address the causes behind contradictory theoretical reports as well as discrepancies between theoretical and experimental findings. We suggest that the most likely ferroelectric phase is the $Pca2_1$ phase, and its ``rhombohedrally distorted" state could also explain experimental indications of a rhombohedral symmetry. 

A major challenge in fabricating ferroelectric hafnia-based thin films is to reduce the volume fractions of nonpolar phases while promoting the presence of the desired ferroelectric phases. In the section `Stabilization mechanisms of ferroelectric phases', we will discuss various extrinsic factors suggested theoretically to contribute to the stabilization, either thermodynamically or kinetically, of ferroelectric phases. Due to the intricacies and high computational costs of modeling the effects of dopants and oxygen vacancies, results from computational studies, primarily based on first-principles methods like density functional theory (DFT), are indicative but not definitive. Nevertheless, they strongly suggest that positively charged oxygen vacancies, either introduced by acceptor dopants or formed during the growth process, have a profound impact on the polymorphism kinetics. In the section `Ferroelectric switching', we will discuss the main challenges limiting the applications of ferroelectric hafnia, notably the much higher coercive fields compared to perovskite ferroelectrics. DFT-based investigations have uncovered multiple competing pathways for domain switching at the unit cell level, as well as the motions of domain walls. We will clarify some subtleties and address common misconceptions about studying ferroelectric switching with DFT. The section `Modeling beyond DFT' explores promising computational tools that could broaden our understanding of the dynamics of ferroelectric switching over larger time and length scales. The review concludes with a discussion of several directions for future computational modeling of ferroelectric hafnia. 

\section{DIVERSE STRUCTURAL POLYMORPHS}

\subsection{Nonpolar and polar phases}

\begin{figure}[t]
\includegraphics[width=1.0 \textwidth]{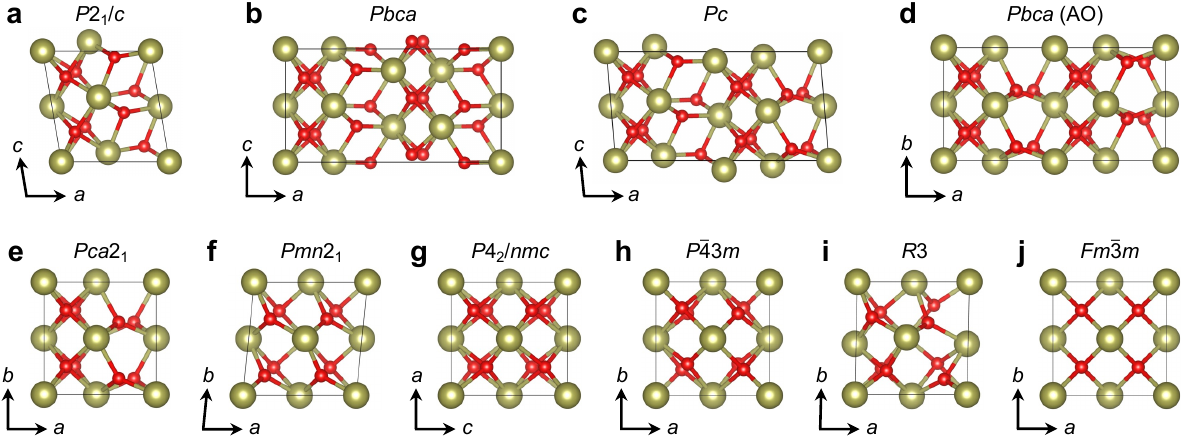}
\caption{Atomic structures of nonpolar and polar HfO$_2$ phases. The Hf and O atoms are denoted by golden and red spheres, respectively. (\textbf{a}) Nonpolar monoclinic $P2_1/c$. (\textbf{b}) Nonpolar orthorhombic $Pbca$. (\textbf{c}) Polar monoclinic $Pc$. (\textbf{d}) Antipolar orthorhombic ($AO$) $Pbca$. (\textbf{e}) Polar orthorhombic $Pca2_1$. (\textbf{f}) Polar orthorhombic $Pmn2_1$. (\textbf{g}) Nonpolar tetragonal $P4_2/nmc$. (\textbf{h}) Nonpolar cubic $P\bar{4}3m$. (\textbf{i}) Polar rhombohedral $R3$. (\textbf{j}) Nonpolar cubic $Fm\bar{3}m$.}
\label{fig_phase}
\end{figure}

The fluorite-structured HfO$_2$ is known to form many polymorphs despite its simple chemical composition (Figure~\ref{fig_phase}). For bulk HfO$_2$, the most stable polymorph at room temperature and ambient pressure is the monoclinic ($M$) $P2_1/c$ phase~\cite{Ohtaka01p1369}. This stable polymorph transforms to the tetragonal ($T$) $P4_2/nmc$ phase, and then to the cubic $Fm\bar{3}m$ phase with increasing temperature. Under increasing pressure, the $M$ phase transitions to the antipolar orthorhombic ($AO$) $Pbca$ phase. Given that all these polymorphs are nonpolar, the discovery of ferroelectricity in hafnia-based thin films was a surprise. The origin of the unexpected ferroelectricity has been attributed to the formation of four polar polymorphs, including two orthorhombic $Pca2_1$ and $Pmn2_1$ phases, and two rhombohedral $R3m$ and $R3$ phases.

The $T$ phase is commonly regarded as a precursor for the formation of low-temperature phases like the nonpolar $M$ phase and the polar $Pca2_1$ due to their low transition barriers~\cite{Ma23p096801}. Unlike the high-symmetry $T$ phase, the structures of the $M$ and $Pca2_1$ phases (Figure~\ref{fig_phase}\textbf{a} and \textbf{e}) consist of alternating fourfold-coordinated (4C) and threefold-coordinated (3C) oxygen atoms. Specifically, the asymmetric displacements of the 3C oxygen atoms in $Pca2_1$ break the inversion symmetry and are often believed to be responsible for the emergence of the macroscopic spontaneous polarization~\cite{Yuan23p94}. However, as we will discuss below (see the section `Ferroelectric switching'), both 4C and 3C oxygen atoms may be involved in the switching process and contribute to the switching current measured in experiments. Within the energy window between the ground-state $M$ phase and the metastable $Pca2_1$ phase, two distinct low-energy orthorhombic $Pbca$ phases~\cite{Kersch21p2100074} can be identified, the nonpolar $Pbca$ and the anitpolar $Pbca$ (Figure~\ref{fig_phase}\textbf{b} and \textbf{d}), which can be viewed as a mirror-reflected arrangements of $M$ and $Pca2_1$ phases, respectively. Note that a recent DFT study predicted a new polar monoclinic $Pc$ phase as another mirror-reflected interphase (Figure~\ref{fig_phase}\textbf{c}), with an energy lower than that of the polar $Pca2_1$ phase~\cite{Antunes22p2100636}. This prediction was supported by indicative evidence from images obtained through scanning transmission electron microscopy~\cite{Du21p986}. Another polar orthorhombic phase, $Pmn2_1$, with energy higher than $Pca2_1$, has been theoretically identified by deforming the $T$ phase using a DFT-based structure search method~\cite{Huan14p064111}. The calculated spontaneous polarization of $Pca2_1$ and $Pmn2_1$ phases is 50.3 and 56.2 $\mu$C/cm$^2$, respectively. 

\begin{figure}[t]
\includegraphics[width=0.8 \textwidth]{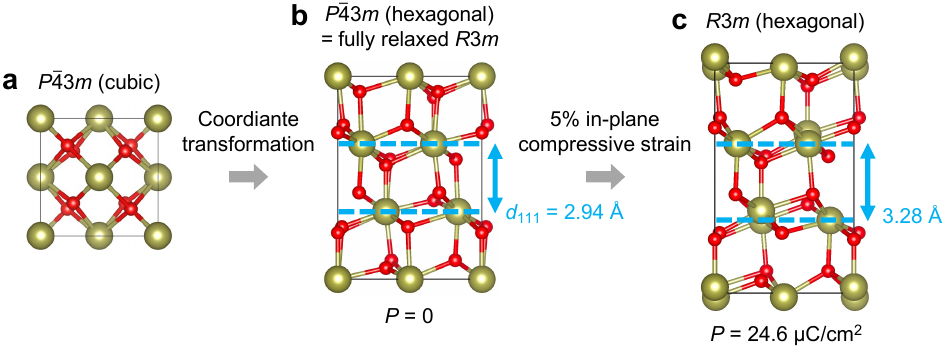}
\caption{Epitaxial strain-induced phase transition from the nonpolar cubic $P\bar{4}3m$ phase to the polar rhombohedral $R3m$ phase. (\textbf{a}) $P\bar{4}3m$ in a 12-atom cubic unit cell. (\textbf{b}) $P\bar{4}3m$ in a 36-atom hexagonal unit cell. Note that the fully relaxed $R3m$ is identical to $P\bar{4}3m$ in a hexgonal unit cell. (\textbf{c}) $R3m$ under 5\% in-plane compressive strain. This strain induces a polarization of 24.6~$\mu$C/cm$^2$, with the out-of-plane interplanar spacing $d_{111}$ expands from 2.94 \AA~to 3.28 \AA.}
\label{fig_r3m}
\end{figure}

Additionally, two high-energy polar rhombohedral phases, $R3m$ and $R3$, have also been proposed~\cite{Wei18p1095}. In DFT calculations, these two rhombohedral phases can be modeled with either a 36-atom hexagonal unit cell or a 12-atom pseudocubic unit cell. The DFT results suggest that the unstrained $R3m$ phase has a negligible polarization of 0.1 $\mu$C/cm$^2$~\cite{Wei18p1095}, indicating that it can be considered nonpolar. This unstrained phase is nearly identical to the nonpolar cubic $P\bar{4}3m$ phase~\cite{Zhu23pL060102}. A polar $R3m$ phase can only be obtained by applying a giant in-plane equibiaxial strain within the (111)-oriented hexagonal supercell of $P\bar{4}3m$ (Table~\ref{tab_phase} and Figure~\ref{fig_r3m}). Notably, even with a 5\% in-plane compressive strain, the polarization is limited to 24.6 $\mu$C/cm$^2$, which is significantly lower than that of the $Pca2_1$ phase. Therefore, based on DFT analysis, we suggest that the $Pca2_1$ phase is intrinsically the most plausible candidate for the ferroelectric phase in hafnia-based thin films, given its thermodynamic stability and polarization value that closely align with experimental results.

\subsection{Characterization of the ferroelectric phase}

Ferroelectricity in hafnia-based thin films has been attributed to the orthorhombic $Pca2_1$ phase since its discovery~\cite{Boscke11p102903, Park15p1811, Shimizu15p032910, Sang15p162905}. Other polar phases, including the orthorhombic $Pmn2_1$~\cite{Huan14p064111, Qi20p257603}, as well as the rhombohedral $R3m$~\cite{Wei18p1095, Nukala21p630, Wang23p558} and $R3$~\cite{Wei18p1095, Begon-Lours20p043401} phases, have also been proposed to account for the observed ferroelectricity in different thin films on various substrates. A broad debate persisted in recent years over the exact ferroelectric phase, particularly between the orthorhombic $Pca2_1$ and rhombohedral $R3m$ phases. Due to the metastable nature of both phases, as-grown thin films often contain nonpolar phases, complicating phase determination in experiments.

The $R3m$ phase was initially suggested by \citeauthor{Wei18p1095} to explain the observed rhombohedral symmetry, characterized by a deviation of the out-of-plane interplanar spacing, $d_{111}$, from the in-plane spacings, $d_{11\bar{1}}$, $d_{1\bar{1}1}$, and $d_{\bar{1}11}$, in Hf$_{0.5}$Zr$_{0.5}$O$_{2}$ (HZO) epitaxial thin films grown on (La, Sr)MnO$_3$ (LSMO) electrode and SrTiO$_3$ (STO) substrate with pulsed laser deposition (PLD)~\cite{Wei18p1095}. Based on the electron microscopy analysis, particularly the observation of forbidden spots in selected area electron diffraction (SAED) patterns, such as (001) and (1$\bar{1}$0) in the [110] zone-axis pattern as well as (100) spot in the [101] zone-axis pattern, the authors suggested that the phase is non-orthorhombic and classified it as rhombohedral. DFT calculations in the same work revealed that inducing substantial polarization in the $R3m$ phase requires a giant in-plane strain, which was confirmed in a recent DFT investigation~\cite{Zhu23pL060102}. Specifically, a $d_{111}$ of 3.28 \AA, corresponding to an in-plane compressive strain of 5\%, can only induce a polarization of 24.6 $\mu$C/cm$^2$, smaller than the experimental value of 34 $\mu$C/cm$^2$. However, the experimental value of $d_{111}$ characterized by X-ray diffraction (XRD) measurements is at most 3.27 \AA~(only in the thinnest film of 1.5~nm). The values of $d_{111}$ reported in subsequent experimental studies of epitaxial ferroelectric hafnia-based films typically range between 2.95 and 3.02 \AA~\cite{Estandia19p1449, Begon-Lours20p043401, Zheng21p172904, Yun22p903}, indicating a lack of large in-plane compressive strain. Moreover, due to the large lattice mismatch between LSMO and HZO, HZO films grown on LSMO/STO adopt the domain-matching epitaxy (DME)~\cite{Fina21p1530, Estandia20p3801}, which often results in small in-plane strains, making it unlikely that an appropriate strain necessary to polarize the $R3m$ phase can be imposed. 

Another polar rhombohedral phase, $R3$, has also been considered to explain the ferroelectricity in epitaxial HZO/LSMO/STO thin films~\cite{Wei18p1095}. In contrast to the $R3m$ phase, the $R3$ phase is inherently polar at its unstrained state (Table~\ref{tab_phase}). \citeauthor{Begon-Lours20p043401} reported rhombohedral symmetry in HZO thin films grown on the GaN(0001)/Si(111) substrate and attributed it to the formation of $R3$ phase~\cite{Begon-Lours20p043401}, although the reported rhombohedral distortion angle (0.6\degree) is still considerably smaller than that (1.3\degree) in the $R3$ phase.

The hypothesis of rhombohedral $R3m$ and $R3$ phases has been questioned due to their high intrinsic energies (Table~\ref{tab_phase}) and the discrepancies between theoretical and experimental results as discussed above~\cite{Qi20p257603, Yun22p903, Zhu23pL060102}. \citeauthor{Qi20p257603} simulated the SAED patterns of various HfO$_2$ polymorphs and found that the pattern of the orthorhombic $Pmn2_1$ phase aligns closely with the experimental observations, suggesting that the kinetically stabilized $Pmn2_1$ could explain the observed ferroelectricity in these films~\cite{Qi20p257603}. \citeauthor{Yun22p903} suggested that the rhombohedral symmetry might not be intrinsic to the polar phase in thin films and could instead arise from the ``rhombohedral distortion" whithin the orthorhombic $Pca2_1$ phase (with a distortion angle of approximately 0.25--0.41\degree)~\cite{Yun22p903}. Their DFT calculations indicated that a distortion angle of 0.4\degree ~only slightly increases the energy of $Pca2_1$ phase by 6 meV per unit cell, suggesting a relatively flat potential energy surface under rhombohedral distortion. A recent DFT high-throughtput (HT) study demonstrated that the $Pca2_1$ phase remains the most stable polar phase across a broad range of epitaxial conditions~\cite{Zhu23pL060102}. Introducing rhombohedral distortion to (111)-oriented $Pca2_1$ can also result in XRD spectra featuring different $d_{111}$, $d_{11\bar{1}}$, $d_{1\bar{1}1}$, and $d_{\bar{1}11}$ spacings. Nevertheless, given the complexity in polycrystalline thin films, we do not rule out the emergence of polar rhombohedral phases due to extrinsic factors.

\subsection{Electron microscopy analysis with computational modeling}

\begin{figure}[t]
\includegraphics[width=0.9 \textwidth]{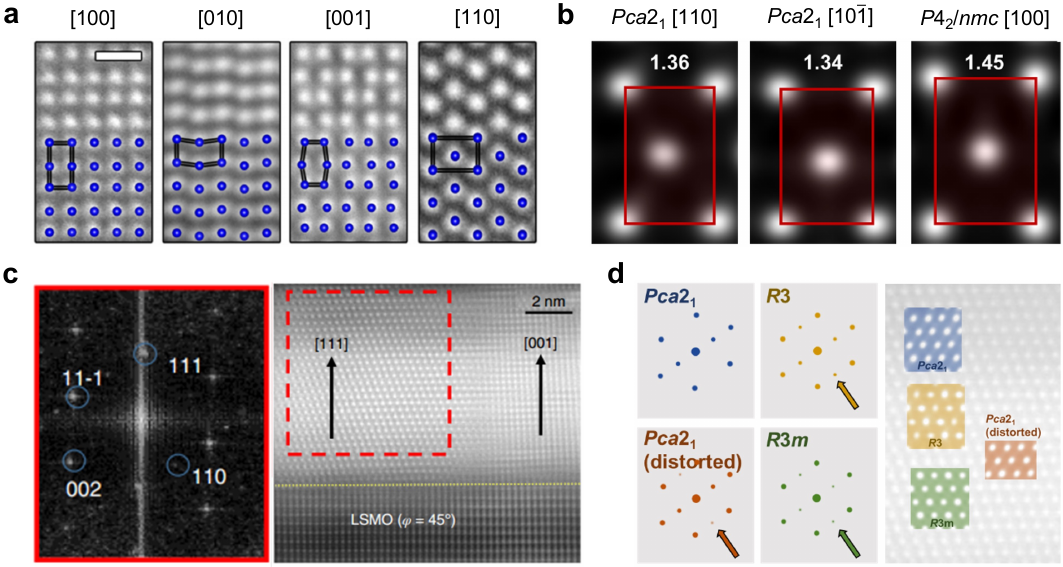}
\caption{Electron microscopy analysis of HfO$_2$. (\textbf{a}) HAADF-STEM images acquired from four different zone axes, overlaid with the Hf sublattice. (\textbf{b}) Simulated HAADF images for $Pca2_1$ and $P4_2/nmc$ phases with Dr. Probe~\cite{Barthel18p1}, illustrating the Hf sublattice. Insets show the aspect ratios of the Hf sublattice rectangle. (\textbf{c}) HAADF-STEM image of the HZO film observed along the [1$\bar{1}$0] zone axis of the film. A fast Fourier transform of the [111] domain is depicted on the left. (\textbf{d}) Simulated electron diffraction patterns for the $Pca2_1$, distorted $Pca2_1$, $R3$, and $R3m$ structures using SingleCrystal. Panels \textbf{a} and \textbf{c} are reproduced with permissions from refs.\cite{Sang15p162905} and~\cite{Wei18p1095}, respectively.}
\label{fig_stem}
\end{figure}

Electron microscopy studies play a pivotal and immensely helpful role in investigating hafnia-based ferroelectrics, particularly in determining their phases and symmetries~\cite{Shiraishi22p118091, Cheng22p645}. In the quest to determine phase and symmetry, direct imaging techniques such as high-angle annular dark-field image (HAADF), annular bright-field image (ABF), and ptychography~\cite{Jiang18p343}, along with SAED and convergent-beam electron diffraction (CBED)~\cite{Kim22p2107620}, are broadly employed. Typically, when the oxygen lattice is visible, the phase or symmetry can be unambiguously determined, although this often requires additional efforts, including stringent criteria for sample thickness and minimal misalignment from the exact zone axis, with even stricter tolerances than those for HAADF imaging~\cite{Gao18p177}. In some cases, employing only HAADF imaging to determine the phase and symmetry is simpler. For instance, \citeauthor{Sang15p162905} utilized atomic-scale HAADF image to analyze atomic positions and lattice parameters, directly identifying the polar $Pca2_1$ phase of HfO$_2$ in thin films for the first time (Figure~\ref{fig_stem}\textbf{a})~\cite{Sang15p162905}. However, when using HAADF imaging alone, extra caution is necessary as projections along different zone axes with varying symmetries may appear quite similar, leading to potential errors. For example, the Hf lattice in the [110] and [10$\bar{1}$] zone axes of the orthorhombic phase closely resembles that of the [100] zone axis of the tetragonal phase. To differentiate these phases, a deeper analysis involving a quantitative examination of the lattice rectangle ratio is essential; for the orthorhombic phase, the ratio is 1.35, whereas for the tetragonal phase, it is 1.45 (Figure~\ref{fig_stem}\textbf{b}). 

Regarding the electron diffraction method, SAED is typically quite useful and straightforward for differentiating symmetries, as different symmetries have distinct diffraction extinction conditions. Nonetheless, it's essential to exercise caution given the dynamic effects observed in electron diffraction patterns, such as the appearance of forbidden spots caused by secondary diffraction~\cite{Carter16TEM}. In the case of HfO$_2$, additional caution is warranted due to the susceptibility of the phase and symmetry to modifications imposed by substrate constraints or local strain states. This is particularly important in polycrystalline and multiphase thin films, where the local structure is more complex~\cite{Lu16p68, Xiao24p3620}. As a result, a variety of phase variants may occur~\cite{Zhu23pL060102}, which may be only slightly distorted from their ideal original symmetries such that certain indicative spots that typically differentiate between phases may lose their reliability. 

Consider the rhombohedral phase as an illustration~\cite{Wei18p1095}. The presence of the (110) spot in the [1$\bar{1}$0] zone-axis pattern, forbidden in the original orthorhombic phase, has historically been taken as determinative evidence for the existence of the rhombohedral phase (Figure~\ref{fig_stem}\textbf{c}). However, our study has revealed that even a slightly distorted orthorhombic phase can produce the (110) spot (Figure~\ref{fig_stem}\textbf{d}). Such a phase has been demonstrated to be more energetically favorable than the original orthorhombic phase and possesses polar characteristics. Given the complexity of symmetries in HfO$_2$, prioritizing CBED over SAED for accurate symmetry determination is advisable. Additionally, a more comprehensive analysis that combines real- and reciprocal-space information is likely to be more effective in accurately characterizing the material's properties.

\subsection{Symmetry mode analysis of $Pca2_1$}

\begin{figure}[t]
\includegraphics[width=1.0 \textwidth]{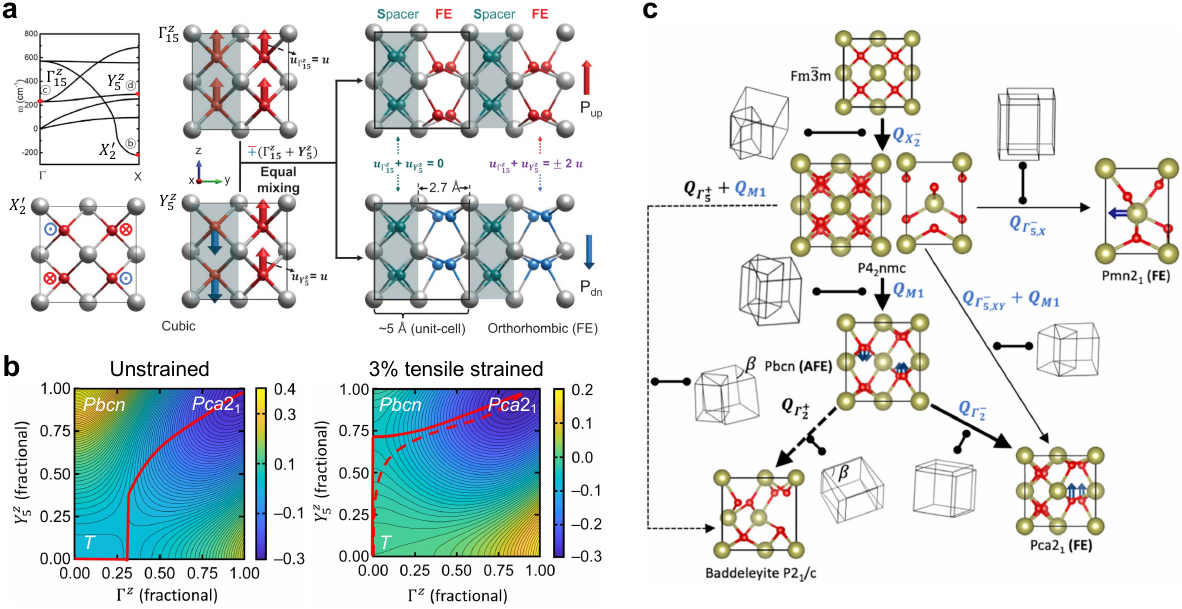}
\caption{Symmetry mode analysis of $Pca2_1$ HfO$_2$. (\textbf{a}) Structural origin of the $Pca2_1$ phase based on the phonon dispersion analysis of the cubic $Fm\bar{3}m$ phase. The condensation of the unstable tetragonal mode ($X'_2$) and stable polar ($\Gamma_{15}^z$) and antipolar ($Y_{5}^z$) modes results in the $Pca2_1$ phase. The equal mixing of the zone-center polar mode $\Gamma_{15}^z$ and zone-boundary antipolar mode $Y_{5}^z$ leads to the alternating layers of four-fold-coordinated oxygen atoms (spacer) and three-fold-coordinated oxygen atoms (FE) in the $Pca2_1$ phase. (\textbf{b}) Strain-induced two-step phase transitions of $T\rightarrow Pbcn \rightarrow Pca2_1$. In the unstrained case, the $Pbcn$ phase is unstable. A 3\% tensile strain along the tetragonal-model direction triggers $T \rightarrow Pbcn$, followed by another transition to $Pca2_1$ due to the coupling between the polar $\Gamma^z$ and antipolar $Y_{5}^z$ modes. (\textbf{c}) Symmetry relationships between different HfO$_2$ phases. The $Pbcn$ phase is the parent phase of $Pca2_1$, related by a soft mode. Panels \textbf{a}, \textbf{b}, and \textbf{c} are reproduced with permissions from refs.~\cite{Lee20p1343}, \cite{Zhou22peadd5953}, and~\cite{Raeliarijaona23p094109}, respectively.}
\label{fig_mode}
\end{figure}

Experimentally, hafnia-based thin films appear to be immune to the depolarization effect~\cite{Cheema20p478} that destabilizes spontaneous polarization in perovskite ferroelectric thin films. \citeauthor{Lee20p1343} proposed that the $Pca2_1$ phase of HfO$_2$ is a material realization of flat phonon bands, resulting in intrinsically localized electric dipoles~\cite{Lee20p1343} that are less affected by domain walls and surface exposure. Additionally, the in-phase condensation of polar $\Gamma_{15}^z$ and anti-polar $Y_5^z$ lattice modes in the cubic phase leads to alternating layers of 4C and 3C oxygen atoms in the polar $Pca2_1$ phase (Figure~\ref{fig_mode}\textbf{a}), suggesting an improper nature of ferroelectricity. \citeauthor{Delodovici21p064405} developed a Landau free energy model using the $Ccce$ phase as the reference, based on DFT calculations. This model revealed that a trilinear coupling between a polar unstable zone-center mode and two stable zone-boundary modes, with nonpolar and polar features, drives the ferroelectric phase transition in HfO$_2$. In comparison, \citeauthor{Zhou22peadd5953} started with the cubic $Fm\bar{3}m$ phase and bridged the pathway connecting $Fm\bar{3}m$ and $Pca2_1$ phases with tetragonal, antipolar, and polar modes~\cite{Zhou22peadd5953}. The high-symmetry $Fm\bar{3}m$ phase transforms to tetragonal $P4_2/nmc$ under the tetragonal-mode deformation. Subsequent inclusion of the antipolar mode leads to an antipolar $Pbcn$ phase. The polar $Pca2_1$ phase is then obtained by further incorporating the polar mode into the $Pbcn$ structure. They suggested that a tensile strain along the tetragonal-mode direction triggers a phase transition from $P4_2/nmc$ to $Pbcn$ (Figure~\ref{fig_mode}\textbf{b}). Moreover, strong cooperative coupling between the polar and antipolar modes facilitates a second phase transition from $Pbcn$ to $Pca2_1$~\cite{Zhou22peadd5953}. \citeauthor{Raeliarijaona23p094109} later pointed out that depending on the choice of parent phase during the symmetry mode analysis~\cite{Huang14p064111,Aramberri23p95,Delodovici21p064405}, ferroelectric HfO$_2$ could be viewed as either an improper or proper ferroelectric. Their conclusion is that ferroelectric hafnia is a proper ferroelectric, as there is a single polar soft mode connecting the parent antipolar $Pbcn$ phase to the polar $Pca2_1$ phase, whereas $P4_2/nmc$ is the parent nonpolar phase for the polar $Pmn2_1$ phase (Figure~\ref{fig_mode}\textbf{c})~\cite{Raeliarijaona23p094109}. More recently, \citeauthor{Aramberri23p95}, assuming a uniaxial ferroic order for HfO$_2$, developed a theoretical framework using the $Pbcm$ phase as the reference state, which connects all low-energy polymorphs such as $M$, $Pbca$, $AO$, and $Pca2_1$ phases through simple phonon distortions~\cite{Aramberri23p95}.

In general, identifying symmetry-adapted lattice modes that contribute to the ferroelectric phase in HfO$_2$ is crucial for establishing appropriate order parameters, which form the foundation for developing perturbation theories, including phenomenological Ginzburg-Landau free-energy models, effective Hamiltonians~\cite{Zhong94p1861}, and second-principles potentials~\cite{Wojdel13p305401}. Developing such computationally efficient thermodynamic models will be helpful in improving our understanding of ferroelectric behavior of HfO$_2$ in a physically transparent manner.

Given the prevalence of several reference phases in the literature, it is natural to wonder which one is the most suitable. The answer to this question is not straightforward, as each reference state offers distinct advantages. For example, the break of the 4-fold axis of the $T$ phase naturally leads to four symmetry-equivalent domains which conveniently explains the experimental observations of 90$^{\circ}$ domain walls~\cite{Shimizu18p212901, Lederer21p202100086, Zhou22p117920}, justifying the choice of the $T$ phase as the high-symmetry reference structure~\cite{Zhou22peadd5953}. However, the $T$ phase lacks polar instability~\cite{Aramberri23p95}, unable to explain the dielectric anomaly upon heating Hf$_{0.5}$Zr$_{0.5}$O$_2$~\cite{Schroeder22p2200265}. Choosing the $Pbcn$~\cite{Raeliarijaona23p094109} or $Pbcm$~\cite{Aramberri23p95} phase as the reference naturally explains this soft-mode-driven proper ferroelectric behavior. The unified description of all low-energy polymorphs starting with the $Pbcm$ phase is evidently appealing~\cite{Aramberri23p95}, although this phase seems irrelevant for the lowest-energy-switching pathway at the unit cell level (see the section `Ferroelectric switching'). We note that our large-scale MD simulations reveal both temperature-driven $Pca2_1\rightarrow Pbcn$ and $Pca2_1 \rightarrow P4_2/nmc$, depending on the composition of hafnia (see the section `Modeling beyond DFT'). This suggests that variations in composition, along with the accompanying strain effects, can influence the nature of ferroelectric phase transitions in hafnia-based ferroelectrics. Therefore, a productive approach would be to combine MD simulations with symmetry mode analysis. The former identifies atomistic mechanisms with reduced human bias, while the latter provides a clearer physical interpretation.

\section{STABILIZATION MECHANISMS OF FERROELECTRIC PHASES}

Unlike conventional perovskite-based ferroelectrics, where polar phases are the ground states below the Curie temperature, the polar phases of HfO$_2$ are all metastable compared to the nonpolar $M$ phase. A fundamental challenge in fabricating hafnia-based ferroelectric thin films is suppressing the formation of the nonpolar $M$ phase while promoting the emergence of the desired ferroelectric phases. Various extrinsic factors, such as oxygen vacancies, dopants, strain, electric fields, surface and interface effects, have been proposed theoretically to contribute to the stabilization of ferroelectric phases, either thermodynamically or kinetically.

\subsection{Surface and interface}

Experimental studies have shown that hafnia-based ferroelectrics exhibit a \textit{reverse} size effect, where polarization increases as film thickness decreases~\cite{HyukPark13p242905, Cheema20p478}. This unusual thickness-dependence of ferroelectricity in hafnia-based thin films has been attributed to the surface energy effect~\cite{Cheema20p478}. \citeauthor{Materlik15p134109} built a free energy model that includes the surface energy term ($\gamma$), indicating that the ferroelectric $Pca2_1$ phase can be stabilized at a grain size of 3$-$5 nm over the $M$ phase~\cite{Materlik15p134109}. However, later comprehensive surface energy calculations considering major crystallographic planes revealed that the polar $Pca2_1$ phase possesses higher surface energy than the nonpolar $M$ and $T$ phases ($\gamma_{O} > \gamma_{M} > \gamma_{T}$, see Table~\ref{tab_surface})~\cite{Batra16p172902, Wu20p252904, YoungLee23p074102}. \citeauthor{Batra16p172902} reported that the polar $Pmn2_1$ phase has the lowest surface energy in (001) orientation and proposed that the effect of surface energy could stabilize the polar $Pmn2_1$ phase and the nonpolar precursor $T$ phase, which can then transform into the polar $Pca2_1$ phase~\cite{Batra16p172902}. It was suggested that concerted effects of surface energy and dopants can thermodynamically stabilize the $T$ phase, followed by a transition into the polar $Pca2_1$ phase during the thermal annealing process~\cite{Wu20p252904}. The surface effect has also been used to explain the stabilization of the polar rhombohedral $R3m$ phase in ultrathin films~\cite{Zhang20p014068, ElBoutaybi22p074406}. We emphasize that the calculated surface energies strongly depend on the termination, stoichiometry, and polarization direction of the slab models (Table~\ref{tab_surface})~\cite{Chen08p46, Acosta21p124417, Acosta23p124401}. The difficulty to obtain reliable surface energies makes it challenging to fully understand the surface effect and could lead to inconsistencies between theory and experiments~\cite{Park17p9973}. Comprehensive and reliable DFT calculations of surface energies for different polymorphs of HfO$_2$ will be crucial to addressing this key question. Additionally, a quantitative understanding of the interface energy for the phase interface separating different polymorphs~\cite{Falkowski21p032905, YoungLee23p074102} would be useful in constructing a nucleation-based growth model to analyze phase transition kinetics during the annealing process in film fabrication.

The charge transfer between the metal electrode and hafnia plays an unexpected role in stabilizing the ferroelectric phase, as highlighted in recent experimental findings. \citeauthor{Shi23p1780} observed enhanced ferroelectricity in HZO thin films grown on MnO$_2$-terminated LSMO by intentionally engineering the HZO/LSMO interface~\cite{Shi23p1780}. This enhancement was attributed to hole doping from MnO$_2$-terminated LSMO into hafnia, which thermodynamically promotes the formation of the polar $Pca2_1$ phase over the nonpolar $M$ phase. Currently, theoretical studies on interface effects, particularly the interplay between charge transfer and structural polymorphism, remain limited, partly because large-scale interface models are needed to simulate experimental domain-matching systems.

\subsection{Epitaxial strain}

Strain engineering of ferroelectric perovskites through thin-film epitaxy has significantly advanced the understanding of ferroelectric physics and led to the discovery of novel topological polar structures~\cite{Tang15p547,Yadav16p198,Das19p368}. By leveraging the lattice mismatch between ferroelectric perovskites and their substrates, various degrees of freedom such as spin, charge, and domain can be tailored to realize new functionalities~\cite{Ramesh19p257}. Epitaxial strain is often cited as an important factor for stabilizing the metastable $Pca2_1$ phase in hafnia thin films, but the situation is more complex than it might seem.

Experimentally, hafnia-based thin films grown on different substrates can adopt two distinct types of epitaxy. One is the conventional lattice-matching epitaxy (LME), also known as cube-on-cube epitaxy, where one film lattice coherently matches one substrate lattice. This type of epitaxy is likely to be observed in hafnia films grown on substrates with similar lattice parameters, such as yttria-stabilized zirconia (YSZ) and indium-tin oxide (ITO)~\cite{Shimizu15p032910, Katayama16p112901, Torrejon18p013401}. The other type is the domain-matching epitaxy (DME), where $m$ lattices of film match $n$ lattices of substrate. The DME-type interface is usually found in films grown on largely mismatched substrates, such as (La,Sr)MnO$_3$ (LSMO). The formation of $m$/$n$ domains effectively reduces the actual lattice mismatch and relieves the epitaxial stress~\cite{Fina21p1530, Estandia20p3801}. In this regard, hafnia in thin films with DME might not experience significant straining.

The effects of equibiaxial strain on the thermodynamic ordering of hafnia polymorphs have been extensively studied. In these DFT-based investigations, the equibiaxial strain is applied through scaling the lattice vectors by the same percentage while conserving the in-plane lattice angle. \citeauthor{Batra17p4139} found that the in-plane equibiaxial compressive strain above $\sim$2.5~\% will make $Pca2_1$ with in-plane polarization more stable than $M$, but the antipolar $Pbca$ phase is always lower in energy than $Pca2_1$~\cite{Batra17p4139}. \citeauthor{Fan20p23LT01} reached a similar conclusion for compressive strains and found that a tensile strain larger than 3\% could make $Pca2_1$ more stable than the antipolar $Pbca$ phase, but it would still be higher in energy compared to the nonpolar $M$ phase~\cite{Fan20p23LT01}. These results imply that equibiaxial strain alone cannot make (001)-oriented $Pca2_1$ the most stable compared to the $M$ or $Pbca$ phases.

\citeauthor{Liu19p054404} were the first to theoretically examine the collective effects of growth orientations and equibiaxial strains on the phase stability of HfO$_2$~\cite{Liu19p054404}. Based on the phase transition barriers computed with variable-cell nudged elastic band (NEB) technique, it was suggested that the transformation from the $T$ phase to the $Pca2_1$ phase is kinetically faster than to the antipolar $Pbca$ phase, which shifts the focus to the thermodynamic competition between the $Pca2_1$ and $M$ phases. They reported that the (111)-oriented $Pca2_1$ is thermodynamically more stable than the nonpolar $M$ phase under in-plane compressive strain~\cite{Liu19p054404}. In contrast, \citeauthor{Chae20p054101} reached a different conclusion, reporting that the in-plane tensile strain ($>$4.2\%) favors the $Pca2_1$ phase over the $M$ phase~\cite{Chae20p054101}. These conflicting conclusions stemmed from the fact that the $M$ phase has symmetrically non-equivalent (111) and (11$\bar{1}$) planes, while the former study did not consider the (11$\bar{1}$)-orientated $M$ phase. It is worth noting that when all \{111\} crystallographic orientations are taken into account, the $M$ phase remains the most stable under any in-plane equibiaxial strain conditions~\cite{Zhang20p014068}.

The combined effects of strain and other extrinsic factors have been proposed to enhance the stabilization of the ferroelectric phase. \citeauthor{Batra17p4139} suggested that the combination of in-plane equibiaxial compressive strain and an in-plane electric field can stabilize the $Pca2_1$ phase as the ground state~\cite{Batra17p4139}, although the electric field required is relatively large ($\approx$2 MV/cm). \citeauthor{Bai23p219} introduced a coupling effect of the uniaxial strain, electric field, and oxygen vacancies, suggesting that this mechanical-electrical-chemical coupling can promote the stabilization of the $Pca2_1$ phase~\cite{Bai23p219}. However, it remains unclear how to achieve the necessary multi-factor control in experimental conditions, especially during the film growth process.

The impact of equibiaxial strain on other polar phases such as $R3m$ and $Pmn2_1$ has been examined. As discussed earlier, the polar $R3m$ phase is induced by applying a (111) in-plane equibiaxial compressive strain~\cite{Wei18p1095, Zhang20p014068, Zhu23pL060102} to an otherwise nonpolar $P\bar{4}3m$ phase~\cite{Zhu23pL060102}. However, the high intrinsic energy and large strain required to induce polarization present obstacles to stabilizing this phase. \citeauthor{Zhang20p014068} argued that a concerted effect of epitaxial strain and surface energy could favor the $R3m$ phase in ultrathin films~\cite{Zhang20p014068}. Another study compared the relative stability of $R3m$ and $M$ phases on a ZnO(0001) substrate, indicating that trigonal symmetry imposed by this substrate might facilitate the stabilization of the $R3m$ phase~\cite{Zheng21p172904}, though a distorted $Pca2_1$ phase remains a possibility. Meanwhile, \citeauthor{Qi20p257603} found that a (111) in-plane shear strain could drive $T\rightarrow Pmn2_1$ phase transition, proposing the kinetically stabilized $Pmn2_1$ as an alternate cause of ferroelectricity in films grown on the LSMO/STO(001) substrate~\cite{Qi20p257603}. 

\begin{figure}[t]
\includegraphics[width=1.0 \textwidth]{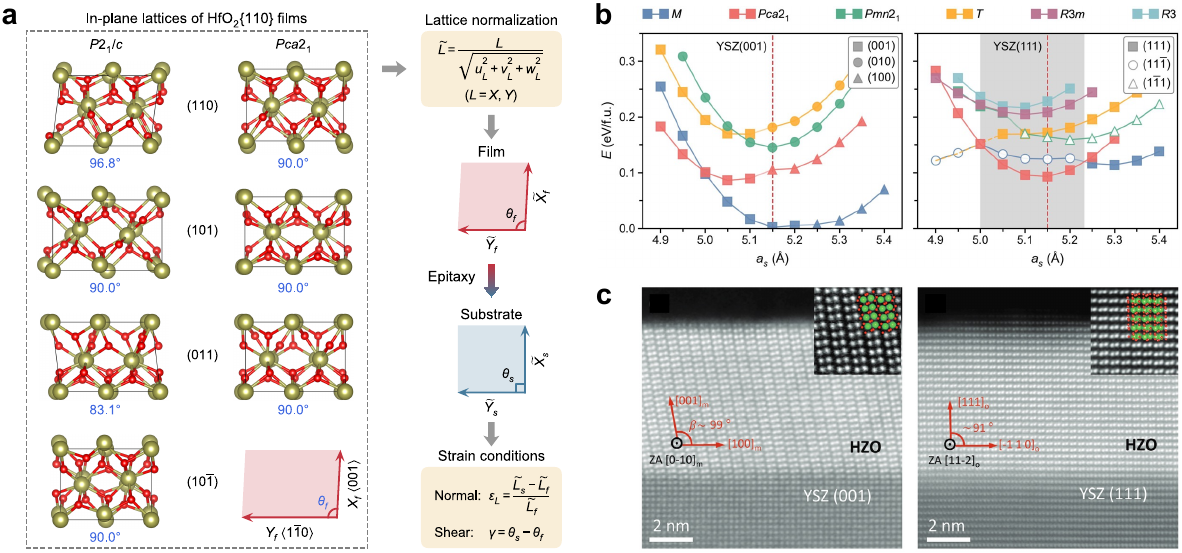}
\caption{Epitaxial strain effect on the phase stability of HfO$_2$. (\textbf{a}) Epitaxial matching of various HfO$_2$ $\left\{110\right\}$ films with a generic substrate and consequent strain conditions. The left panel displays the in-plane lattices for four $P2_1/c$ and three $Pca2_1$ unique growth orientations, each characterized by a distinct set of lattice parameters ($X_f$, $Y_f$, $\theta_f$). By normalizing these in-plane lattice parameters as ($\widetilde{X}_f$, $\widetilde{Y}_f$, $\theta_f$), the strain conditions ($\varepsilon_X$, $\varepsilon_Y$, $\gamma$) of these films imposed by a generic substrate ($\widetilde{X}_s$, $\widetilde{Y}_s$, $\theta_s$) depend on both growth orientation and crystal symmetry. Both normal and shear strains may occur due to the mismatch between the substrate and the film. (\textbf{b}) Thermodynamic stability of HfO$_2$ films under isotropic epitaxial conditions enabled by LME growth in $\left\{001\right\}$ and $\left\{111\right\}$ orientations. (\textbf{c}) Experimental confirmation of selective growth of $M$ and $Pca2_1$ phases in HZO epitaxial thin films deposited on (001)- and (111)-oriented YSZ substrates. Panels \textbf{a} and \textbf{b} are reproduced with permission from ref.~\cite{Zhu23pL060102}. Panel \textbf{c} is reproduced with permission from ref.~\cite{Barriuso24p2300522}.}
\label{fig_strain}
\end{figure}

These DFT-based studies have mainly focused on the equibiaxial strain effect, implying strain alone is insufficient to stabilize the ferroelectric $Pca2_1$ phase as the ground state. However, the epitaxial matching between hafnia-based thin films and their substrates can result in more complex strain conditions, involving both normal (not necessarily equibiaxial) and shear strains due to the low symmetries of relevant HfO$_2$ phases~\cite{Zhu23pL060102}. When growing on a specific substrate, the exact strain conditions depend not only on the matching plane but also on the crystallized phase. As shown in Figure~\ref{fig_strain}\textbf{a}, in \{110\}-oriented thin films, four distinct growth orientations are available for $M$, but only three for $Pca2_1$; their strain conditions for each when grown on the same substrate will be drastically different. Additionally, different epitaxy types, LME and DME, can lead to varying strain conditions. The high-throughput (HT) DFT study by \citeauthor{Zhu23pL060102} systematically evaluated the thermodynamic stability of disparate crystallographic planes and various nonpolar/polar phases under differing epitaxial conditions, showing that shear strain due to orthogonal in-plane lattice constraints imposed by the substrate could thermodynamically favor the polar $Pca2_1$ phase in $\left\{110\right\}$ and (111) growth orientations over other polar forms ($Pmn2_1$, $R3m$, and $R3$), and even nonpolar $M$ phase~\cite{Zhu23pL060102}. These DFT predictions (Figure~\ref{fig_strain}\textbf{b}) are confirmed by a recent experimental study where \citeauthor{Barriuso24p2300522} achieved selective growth of $M$ and $Pca2_1$ phases of HZO on (001)- and (111)-oriented YSZ substrates (Figure~\ref{fig_strain}\textbf{c}), respectively~\cite{Barriuso24p2300522}. It should be noted that these DFT investigations utilized the single-domain approximation to reduce computational costs, overlooking the effects of domain structures on thermodynamic stability. Moreover, while the strain effect on the ferroelectric properties of hafnia- and ziconia-based thin films has been clearly demonstrated~\cite{Kao22p3897,Xu24p2311825}, there is limited theoretical research on the impact of strain on the kinetic barriers of structural polymorphisim. Developing phase-field methods based on accurate thermodynamic potentials of hafnia polymorphs will enhance our understanding of how domain structures and phase mixtures influence the thermodynamic stability of hafnia thin films.

\subsection{Oxygen vacancies}

Long before the discovery of ferroelectricity in HfO$_2$ thin film, oxygen-deficient hafnia (HfO$_{2-x}$) garnered attention for its potential applications in non-volatile resistive random-access memory devices, where the reversible formation and disruption of conductive filaments composed of chain-like oxygen vacancies were key. Given the significance of oxygen vacancies in HfO$_2$, it is not entirely unexpected that they also play a crucial role in ferroelectric hafnia as well. 

Indeed, many experimental studies have emphasized the significant impact of oxygen vacancies on the ferroelectric properties of hafnia-based thin films. \citeauthor{Pal17p022903} found that reducing the oxidant dose during atomic layer deposition (ALD) could suppress the formation of the nonpolar $M$ phase, resulting in up to sixfold improvement in remnant polarization in sub-10 nm HfO$_2$ films without dopants~\cite{Pal17p022903}. \citeauthor{Materano20p3618} also reported that high oxygen contents promote the stabilization the $M$ phase, while short ozone dose times favor the formation of the $T$ and ferroelectric phases in thin films deposited by ALD and physical vapor deposition (PVD)~\cite{Materano20p3618}. \citeauthor{Nukala21p630} emphasized oxygen vacancies as the electrochemical origin of ferroelectricity in hafnia-based compounds by demonstrating reversible migrations of oxygen vacancies and the associated phase transitions employing electron microscopy imaging~\cite{Nukala21p630}. Other phenomena, such as wake-up and fluid imprint, have also been attributed to the redistribution of oxygen vacancies~\cite{Lomenzo15p134105, Buragohain18p222901}.

\begin{figure}[t]
\includegraphics[width=1.0 \textwidth]{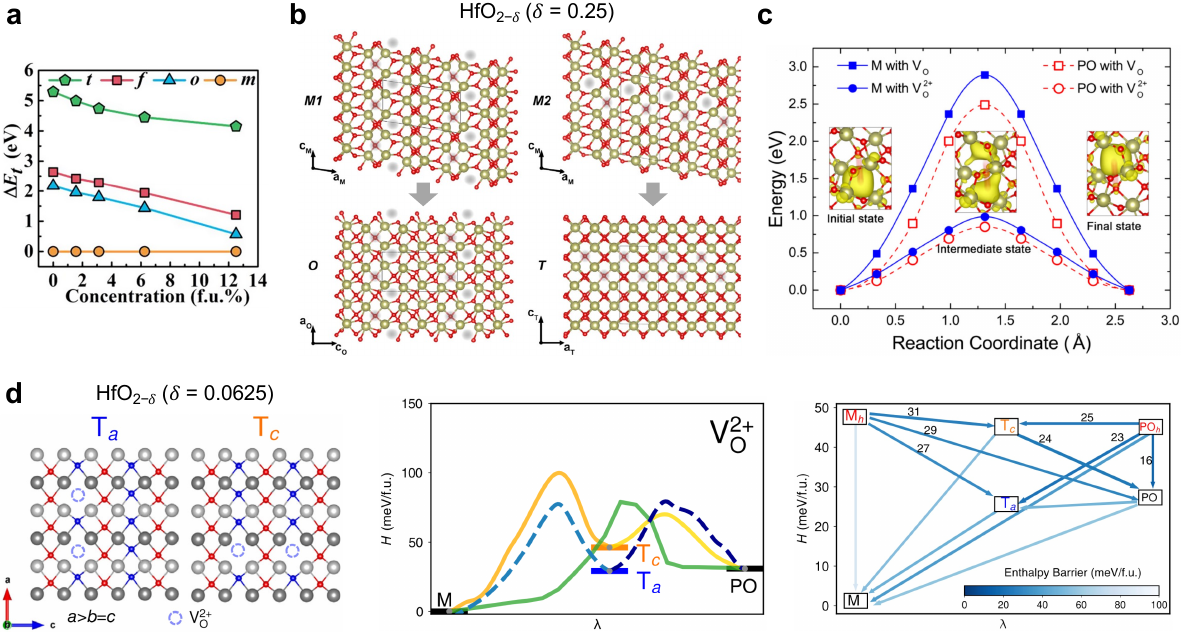}
\caption{Effect of oxygen vacancies on the origin of ferroelectricity in HfO$_2$. (\textbf{a}) Impact of charge-neutral oxygen vacancies (\Vo) on phase stability. A high \Vo~concentration of 12.5 f.u.\% is not sufficient to make the ferroelectric phase ($f$) more stable than the monoclinic phase ($m$). (\textbf{b}) Phase transitions starting from two configurations of the $M$ phase, each at a \Vo~concentration of 12.5 f.u.\%, with different \Vo~orderings. (\textbf{c}) Impacts of the charge state on the diffusion barriers of oxygen vacancy. (\textbf{d}) \Vp-promoted structural polymorphism kinetics. Two different configurations of the $T$ phase, with a pair of \Vp~aligned along the $a$-axis and $c$-axis (denoted as $T_a$ and $T_c$), show drastically different thermodynamic stability and polymorphism kinetics. The presence of \Vp~promotes $T\rightarrow Pca2_1$ (PO) and suppresses $T\rightarrow M$. The phase transition network involving multiple polymorphs of HfO2 containing \Vp~indicates \Vp~can activate both nonpolar-to-polar and polar-to-nonpolar phase transitions. Panels \textbf{a}, \textbf{b}, \textbf{c}, and \textbf{d} are reproduced with permissions from refs.~\cite{Zhou19p143}, \cite{Rushchanskii21p087602}, \cite{He21pL180102}, and~\cite{Ma23p096801}, respectively.}
\label{fig_vo}
\end{figure}

A typical approach for modeling an oxygen vacancy in DFT calculations is to remove an oxygen atom from a supercell of HfO$_2$, effectively creating a charge-neutral oxygen vacancy (\Vo). Due to the high computational cost, most DFT investigations used supercells with no more than 216 oxygen atoms, which restricts modeling to relatively high concentrations of \Vo. \citeauthor{Zhou19p143} found that an increasing concentration of \Vo~tends to reduce the energy difference between the $Pca2_1$ and $M$ phases~\cite{Zhou19p143}. However, even at a high \Vo~concentration of 12.5~f.u.\%, the $M$ phase remains to be much more stable than the $Pca2_1$ phase by 18 meV per unit cell (Figure~\ref{fig_vo}\textbf{a}). \citeauthor{Rushchanskii21p087602} considered the role of \Vo~in the phase transition between $M$ and $Pca2_1$ phases. Using a supercell model of oxygen-deficient HfO$_{2-{\delta}}$ with $\delta=0.25$, they found that oxygen vacancies form two-dimensional extended defects based on a DFT-driven evolutionary algorithm. Interestingly, the two lowest-energy configurations of the $M$ phase exhibit distinct transition behaviors: the lowest-energy configuration kinetically favors the transition to the ferroelectric $Pca2_1$ phase, whereas the second-lowest-energy configuration tends to become the $T$ phase (Figure~\ref{fig_vo}\textbf{b}). But again, the \Vo~concentration in this model reached 12.5 f.u.\%, much higher than typical experimental values of 2–3 f.u.\%~\cite{Pal17p022903, Islamov19p47}. \citeauthor{Materano20p3618} proposed a model in which oxygen vacancies and interstitial can modulate the transition barrier from $T$ to $Pca2_1$ and $M$ phase during crystallization~\cite{Materano20p3618}, but the impact is not very significant. The weak \Vo~effect predicted by DFT seems inconsistent with various experimental evidence indicating a significant impact of oxygen vacancies.

It was later realized that the charge state of oxygen vacancy could reconcile this experimental-theoretical discrepancy. \citeauthor{He21pL180102} computed the diffusion barriers for \Vo~and the doubly positively charged oxygen vacancy (\Vp)~\cite{He21pL180102}. Given that HfO$_2$ has a large band gap, an intuitive understanding of the electronic structure of \Vo~is that it represents two electrons localized at the vacancy site. Further extraction of those two electrons (equivalent to hole doping) leads to \Vp. It was found that \Vo~exhibits a large diffusion barrier of $>2.4$ eV in both $M$ and $Pca2_1$ phases, making it nearly immobile. In contrast, \Vp~has a much lower diffusion barrier, below 1 eV (Figure~\ref{fig_vo}\textbf{c}), suggesting that these charged oxygen vacancies are likely responsible for the observed reversible oxygen vacancy migrations in experiments~\cite{Nukala21p630}. It was reported that beyond a critical concentration of \Vp, the $Pca2_1$ phase could be thermodynamically more stable than the $M$ phase~\cite{He21pL180102}. Subsequent comprehensive calculations of different configurations of $M$ and $Pca2_1$ phases, however, demonstrated that the lowest-energy configuration of the $M$ phase remains more favorable than the lowest-energy configuration of the $Pca2_1$ phase~\cite{Ma23p096801}. Although the presence of \Vp~does not reverse the energetic ordering, \citeauthor{Ma23p096801} discovered that even at a low concentration of 3.125 f.u.\%, local ordering of \Vp~can drastically modulate the relative configurational stability of the same phase. For example, a configuration of the $T$ phase becomes highly unstable when \Vo~becomes \Vp (due to electron entrapment), and its transformation to the polar $Pca2_1$ phase has a lower kinetic barrier of 24 meV/f.u., compared to 54 meV/f.u. for the $T\rightarrow M$ transition. By mapping out the phase transition network involving multiple polymorphs of HfO$_2$ containing \Vp, it was suggested that these charged vacancies act like ``catalysts", enabling various previously forbidden transitions among polar and nonpolar phases (Figure~\ref{fig_vo}\textbf{d}). The robust structural polymorphism kinetics promoted by \Vp~could explain the emergence of polar phases in hafnia thin films, as well as wake-up effects, fluid imprinting, and inertial switching. The significance of the charge state of oxygen vacancy underscores the importance of quantifying their formation energies. Several investigations~\cite{Wei21p2104913,Alam21p084102, Perevalov18p194001} have demonstrated that as the Fermi level decreases, the formation energy of \Vp~in hafnia becomes negative.

The necessity of \Vp~for promoting the formation of the $Pca2_1$ phase aligns with a substantial body of experimental data indicating that nearly all cation dopants inducing ferroelectricity in HfO$_2$-based thin films are $p$-type. This is because substituting Hf$^{4+}$ with acceptor dopants such as Y$^{3+}$ naturally leads to the formation of \Vp~to maintain charge neutrality. The confirmation of intrinsic ferroelectricity in Y-doped HfO$_2$ thin films, which exhibit a large remnant polarization of 64~$\mu$C/cm$^2$, serves as an excellent example~\cite{Yun22p903}. Similarly, the emergence of ferroelectricity in N-doped HfO$_2$~\cite{Xu16p091501} is likely due to \Vp, which compensates for {N$_{\rm O}^-$}. Additionally, the mechanism of \Vp-promoted nonpolar-polar structural polymorphism helps explain why light ion bombardment significantly enhances ferroelectricity in HfO$_2$-based thin films~\cite{Kang22p731}.

We would like to offer a few general comments regarding the subtleties of DFT modeling of oxygen vacancies. First, the charge state of oxygen vacancy is controlled by adjusting the number of electrons in the supercell, with the background charge method employed to maintain the charge neutrality of the whole system when modeling \Vp. A more realistic approach to introducing \Vp~is to replace Hf atoms with divalent or trivalent dopants such as Ca$^{2+}$ and La$^{3+}$, and then compensate for the charge by removing oxygen atoms. Typical acceptor-centered defect complexes are $[{\rm (Ca_{Hf}^{''}}-V_{\rm O}^{\cdot\cdot})^{\times}]$ and $[{\rm (2La_{Hf}^{'}}-V_{\rm O}^{\cdot\cdot})^{\times}]$, where $V_{\rm O}^{\cdot\cdot}$ is the Kr\"oger–Vink notation of \Vp. We believe it is worth investigating the interplay between the charge state of oxygen vacancies and polymorphism kinetics with realistic defect pairs. Second, caution is advised when discussing the impacts of oxygen vacancies on the energetic ordering of hafnia polymorphs, due to the large number of possible oxygen-deficient configurations for the same phase. Identifying the lowest-energy configuration of a given phase is highly nontrivial, especially since the size of the configuration space grows exponentially with the supercell size. Lastly, there exist multiple pathways connecting two phases, depending on the choice of atom-to-atom mapping in DFT-based NEB calculations. An important guiding principle for pathway construction is to ensure the pathway conserves the sign of $X_2^-$ mode (see discussions below), as failing to do so could lead to misinterpretation of the impact of oxygen vacancies on phase transition kinetics.

\subsection{Dopants}

Because ferroelectric hafnia was first discovered in Si-doped HfO$_2$, the effects of doping have attracted great attention from the outset. A number of experimental studies have investigated the impact of various substitutional dopants, such as acceptors, donors, and homovalent dopants on the ferroelectric properties of hafnia-based thin films~\cite{Mueller12pN123,Schroeder18p2752,Richter17p1700131,Richter17p1700131}. Most cation dopants inducing ferroelectricity in HfO$_2$-based thin films are acceptors such as La$^{3+}$ and Y$^{3+}$. This aligns with the theory discussed above that \Vp~facilitates nonpolar-to-polar phase transitions, as acceptor dopants naturally introduce \Vp~to maintain charge neutrality. \citeauthor{Yu23p054052} proposed that La$^{3+}$ doping significantly reduces the formation enthalpy of oxygen vacancies by pushing down the Fermi level, leading to a moderate concentration of \Vp~in most of the chemical potential regions~\cite{Yu23p054052}. Likewise, the emergence of ferroelectricity in nitrogen-doped HfO$_2$ may result from \Vp~ compensating for the negative charge of N$_{\rm O}^-$ ~\cite{Xu16p091501}.

\begin{figure}[t]
\includegraphics[width=1.0 \textwidth]{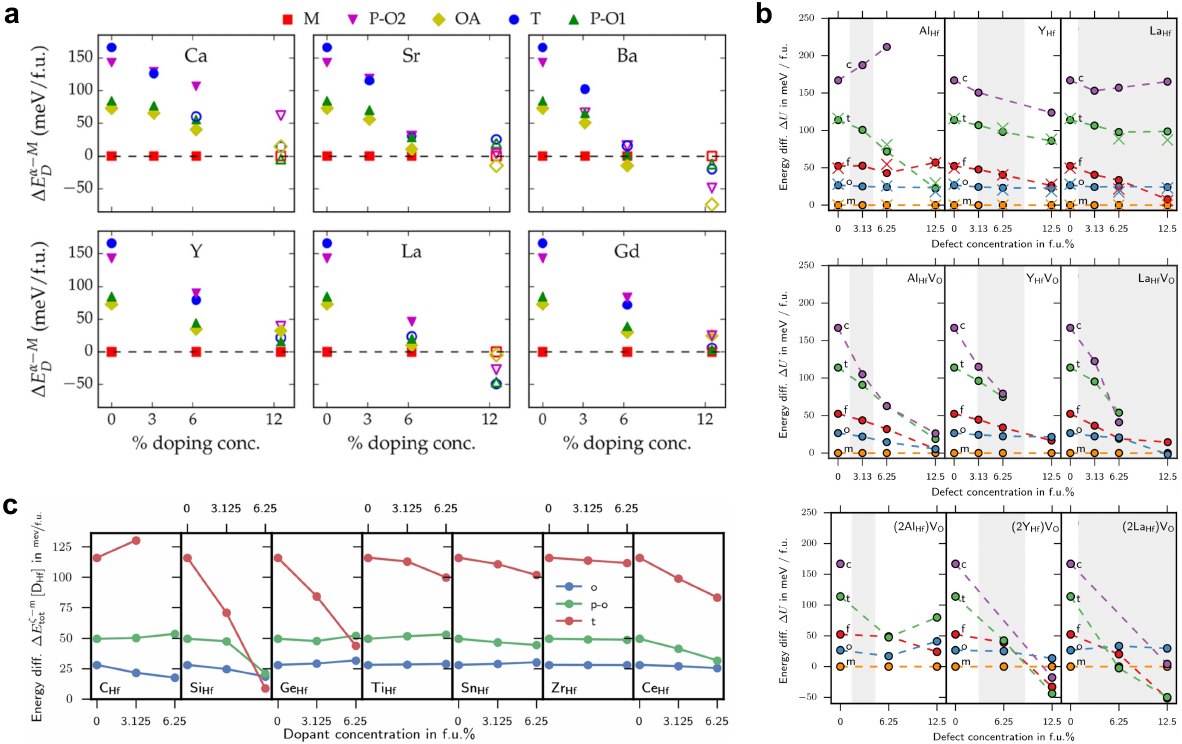}
\caption{Effect of dopants and defect complexes on the phase stability of HfO$_2$. (\textbf{a}) Phase stability in the presence of divalent (Ca, Sr, Ba) and trivalent (Y, La, Gd) dopants with appropriate charge-neutralizing oxygen vacancies. Here, P-O1, P-O2, and OA refer to $Pca2_1$, $Pmn2_1$, and $Pbca$, respectively. (\textbf{b}) Phase stability in the presence of electronically compensated $X_{\rm Hf}$ (top), mixed compensated $X_{\rm Hf}V_{\rm O}$ (middle), and ionically compensated (2$X_{\rm Hf}$)$V_{\rm O}$ (bottom), with $X$ = Al, Y, and La. The labels f and o correspond to $Pca2_1$ and $Pbca$, respectively. (\textbf{c}) Phase stability under four-valent (C, Si, Ge, Ti, Sn, Zr, Ce) doping. The Si doping strongly stabilize the $P4_2/nmc$ phase (t) over $Pbca$ (o) and $Pca2_1$ (p-o) phases. Panels \textbf{a}, \textbf{b}, and \textbf{c} are reproduced with permissions from refs.~\cite{Batra17p9102}, \cite{Materlik18p164101}, and~\cite{Knneth17p254}, respectively.}
\label{fig_dopant}
\end{figure}

\citeauthor{Batra17p9102} examined the effects of nearly 40 different dopants on the stability of hafnia phases based on HT-DFT calculations and a three-stage down-selection strategy. Although none of the dopants make the polar $Pca2_1$ phase the most stable, divalent and trivalent dopants like Ca, Sr, Ba, La, Y, and Gd, along with appropriate charge-neutralizing oxygen vacancies, reduce the energy of the polar phase relative to the $M$ phase considerably (Figure~\ref{fig_dopant}\textbf{a})~\cite{Batra17p9102}. It was inferred that dopants such as Ca, Sr, Ba, and Y, with larger ionic radii and lower electronegativity, are the most effective in stabilizing the polar phase, attributed to the stronger additional bonds formed between the dopant cation and the second-nearest oxygen neighbors. Interestingly, this principle does not fully explain the significance of La as a dopant, which is the most widely used and effective in promoting ferroelectricity in experiments~\cite{Muller13p10.8.1}. \citeauthor{Materlik18p164101} arrived at a somewhat different conclusion when investigating three types of defects associated with trivalent dopants, the electronically compensated $X_{\rm Hf}$, the mixed compensated $X_{\rm Hf}V_{\rm O}$, and the ionically compensated (2$X_{\rm Hf}$)$V_{\rm O}$, with $X$=Al, Y, and La (Figure~\ref{fig_dopant}\textbf{b}). Technically, the modeling of $X_{\rm Hf}$ and $X_{\rm Hf}V_{\rm O}$ employed a combination of adjusting electron numbers and the background charge method to ensure that the oxidation state of $X$ remains +3. This study found that for Y- and La-doping, $X_{\rm Hf}V_{\rm O}$ and (2$X_{\rm Hf}$)$V_{\rm O}$ strongly favor the tetragonal and cubic phases over the ferroelectric $Pca2_1$ phase at a high doping concentration of 12.5 f.u.\%. A similar conclusion was reached for Sr-doped HfO$_2$, where the introduction of vacancies in combination with the Sr dopants destabilizes the ferroelectric phase~\cite{Materlik17p082902}. These findings are not entirely consistent with the theory of \Vp-promoted ferroelectricity~\cite{Ma23p096801}.

Four-valent dopants generally result in robust stabilization of the $T$ phase but only weakly stabilize the $Pca2_1$ phase (Figure~\ref{fig_dopant}\textbf{c})~\cite{Knneth17p254}. Specifically, C, Ge, and Ti barely change the energy difference between the $Pca2_1$ and $M$ phases, while Sn and Ce exhibit a small stabilization effect on the $Pca2_1$ phase. Notably, Si stands out as a unique case, significantly promoting both the $Pca2_1$ and $T$ phases, with a particularly strong impact on the latter~\cite{Fischer08p012908}. The stabilization of the $Pca2_1$ phase by silicon doping has been attributed to the adoption of the more favorable six-fold coordination compared to the less favorable five-fold coordination in the competing $M$ phases~\cite{Knneth17p254}. Nevertheless, a Si doping concentration of up to 12.5 f.u.\% is not enough to favor the $Pca2_1$ over the $M$ phase.

Another four-valent dopant deserving special attention is Zr. Due to the lanthanide contraction, Hf$^{4+}$ and Zr$^{4+}$ have nearly identical ionic radii, leading to similar properties. As a result, HfO$_2$ and ZrO$_2$ are isostructural, undergoing similar temperature- and pressure-driven phase transitions. They can form single-phase solid solutions over the entire composition range~\cite{Mller12p4318, Park17p9973}. Both ZrO$_2$ and HfO$_2$ can adopt the polar $Pca2_1$ phase~\cite{Huang21p116536}, but ZrO$_2$ typically favors the formation of the $T$ phase in thin films, giving rise to antiferroelectric-like polarization-electric field hysteresis loops~\cite{Park15p1811}. The mixed system of Hf$_x$Zr$_{1-x}$O$_2$ (HZO) can support ferroelectricity for a wide range of values of $x$ without further doping. Moreover, HZO thin films could be crystallized at lower temperatures than those based on HfO$_2$, providing an advantage during the integration process~\cite{Mller12p4318}. Despite the similarities between Hf and Zr, explaining why HZO exhibits enhanced ferroelectric properties compared to pure HfO$_2$ remains challenging. This suggests that unique mechanisms could result from the alloying of these two elements, with further research required to understand the underlying causes. Recent experimental investigations have shown that interstitial Hf (Zr) defects can markedly reduce the coercive field of ferroelectric HZO thin films~\cite{Wang23p558}, indicating the need for future theoretical studies on interstitial defects that have been relatively underexplored.

First-principles simulations of the doping effect face challenges similar to those encountered when modeling oxygen vacancies, and often even more so. The interactions among dopants, oxygen vacancies, and their charge states are inherently complex, with computational complexity increasing significantly when considering additional factors like doping concentration, spatial distribution of dopants, and the variety of hafnia polymorphs. For example, HT-DFT calculations of randomly doped HfO$_2$ structures with Si, La, and oxygen vacancies indicated large energy variations among doped configurations due to dopant interactions~\cite{Falkowski18p73}. This hierarchical complexity can quickly lead to computational intractability, which explains why theoretical predictions in literature have not yet reached consistency. A promising solution could be the development of cost-efficient computational tools, such as machine-learning-based force fields, which are capable of simulating doped hafnia systems more effectively. Another important direction for understanding the doping effect is to go beyond the energy-based thermodynamic analysis, shifting the focus on the structural polymorphism kinetics, as no dopant can make the $Pca2_1$ phase the ground state.

Extensive DFT investigations suggest that stabilizing the ferroelectric phase thermodynamically over the competing nonpolar $M$ phase using a single extrinsic factor is quite challenging. One notable exception is that orthogonal in-plane lattice constraints imposed by substrates on \{110\}- and \{111\}-oriented thin films could make the $Pca2_1$ phase most stable thermodynamically , based on DFT investigations assuming a single-domain thin film with LME~\cite{Zhu23pL060102}. The emergence of the metastable ferroelectric phase in epitaxial films with DME and in polycrystalline films with stress-relieving defects, however, remains an open question. Possible methods to realistically address complexities under experimental conditions include MD simulations and phase-field modeling. With accurate potentials for hafnia polymorphs, these techniques would allow for a more effective treatment of interfaces, dopants, and vacancies, as well as the simulation of phase transition kinetics.

\section{FERROELECTRIC SWITCHING}

Understanding the mechanisms of ferroelectric switching in hafnia-based ferroelectrics is important for addressing several critical issues such as high coercive fields and limited endurance, which are the main obstacles hindering their applications. As illustrated in Figure~\ref{fig_phase}\textbf{a}, a striking structural characteristic of the polar $Pca2_1$ phase is the presence of a spacing layer consisting of 4C oxygen atoms that separates 3C oxygen atoms. The 4C and 3C oxygen atoms are commonly referred to as polar oxygen (\Op) and non-polar oxygen (\Onp), respectively. Based on the observation that \Op~and \Onp~layers are arranged in alternating directions perpendicular to the polarization, \citeauthor{Lee20p1343} insightfully pointed out that ferroelectric HfO$_2$ is a material realization of flat phonon bands, where electrical dipoles are highly localized~\cite{Lee20p1343}. This unusual structural feature also hints at the complexity of switching mechanisms for both domains and domain walls.

\subsection{Lattice mode matching and transition pathways}

\begin{figure}[t]
\includegraphics[width=1.0 \textwidth]{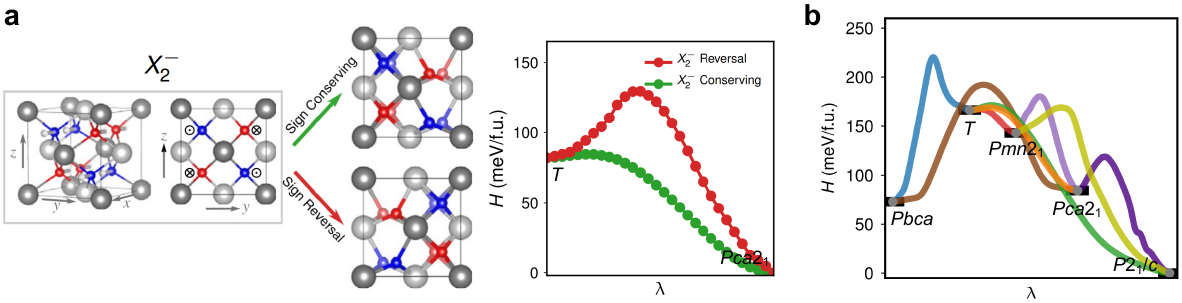}
\caption{$X_2^-$ mode matching and phase transition pathways for HfO$_2$ polymorphs. (\textbf{a}) Schematic illustration of $X_2^-$ mode in the $T$ phase (left) and phase transition pathways between the $T$ and $Pca2_1$ phases with conserved and reversed $X_2^-$ mode (right). (\textbf{b}) Minimum energy pathways between different phases of HfO$_2$ obtained with VCNEB. The transition from the $T$ phase to the antipolar $Pbca$ phase needs to overcome a larger barrier than those for $T\rightarrow Pca2_1$ and $T\rightarrow M$. Reproduced with permission from ref.~\cite{Ma23p096801}.}
\label{fig_transition}
\end{figure}

We start this section by emphasizing the importance of the $X_2^-$ lattice mode, characterized by antiparallel $x$-displacements of neighboring oxygen atoms. The minimum energy paths (MEPs) between two given states of a system are often determined DFT-based NEB method. The performance of NEB calculations requires a predefined atomic mapping scheme, upon which the initial transition path consisting of a discreet set of configurations can be constructed. This is typically achieved by a linear interpolation of the Cartesian coordinates between the initial and final states. Depending on the atom-to-atom mapping chosen, there are multiple possible ways to connect the two states. Employing a mapping that conserves the sign of the $X_2^-$ mode results in a lower enthalpy barrier compared to pathways where this mode is reversed. For instance, the pathway preserving the $X_2^-$ sign between $Pca2_1$ and $P4_2/mmc$ (as shown in Figure~\ref{fig_transition}\textbf{a}) exhibits a substantially lower enthalpy barrier of 2.5 meV per formula unit (f.u.), in contrast to the pathway with the reversed $X_2^-$ mode (48 meV/f.u.). Conserving the sign of $X_2^-$ mode is crucial to the calculating of the energy barrier of both the phase transition and the polarization switching in hafnia~\cite{Qi21parXiv,Ma23p256801}. Figure~\ref{fig_transition}\textbf{b} depicts the MEPs for structural polymorphisms of hafnia, obtained using variable-cell NEB (VCNEB). The main difference between NEB and VCNEB is that VCNEB accommodates changes in lattice constants during solid-solid transformations. This feature enables VCNEB to accurately quantify the intrinsic transition barriers between two phases that have different lattice constants. \citeauthor{Ma23p256801} found that the transition of $T\rightarrow M$, $T\rightarrow Pca2_1$, and $T \rightarrow Pmn2_1$ are all kinetically fast with negligible enthalpy barriers ($<5$~meV/f.u.), whereas the transition of $T \rightarrow Pbca$ requires overcoming a large kinetic barrier despite the antipolar $Pbca$ phase being the second most favored phase thermodynamically. Additionally, transitions from the polar $Pmn2_1$ phase to $M$ and $Pca2_1$ involve relatively small barrier of 28 and 37 meV/f.u., respectively, while a large barrier separates $Pca2_1$ and $M$ phases. These findings suggest that the phase transition kinetics can explain the more common frequent observation of $M$ and $Pca2_1$ phases, rather than $Pmn2_1$ and $Pbca$ phases in as-grown thin films of hafnia. It should be noted that since all these calculations are performed at zero Kelvin, quantifying the finite-temperature free energies of different phases could enhance our understanding of the kinetics of polymorphism and the emergence of ferroelectricity.

\subsection{Unit-cell-level polarization switching}

\begin{figure}[t]
\includegraphics[width=1.0 \textwidth]{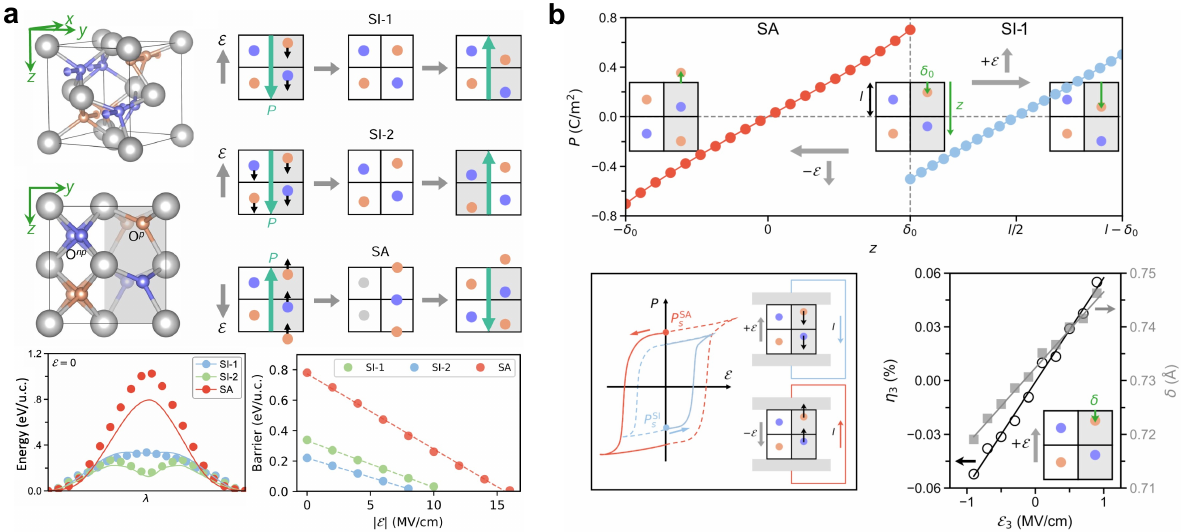}
\caption{Dual-valued remnant polarization and single-valued piezoelectric response in $Pca2_1$ HfO$_2$. (\textbf{a}) Schematics of the $X_2^-$ mode in the $Pca2_1$ phase (left) and shift-inside (SI) and shift-across (SA) switching pathways driven by opposing electric fields $\mathcal{E}$ (right). The bottom panel illustrates the DFT MEPs for SI-1, SI-2, and SA, respectively, alongside the switching barrier plotted against field strength ($|\mathcal{E}|$). We note that in Figure~1e of ref. \cite{Ma23p256801}, which plots barrier versus $|\mathcal{E}|$, a factor of 4 for the field strength was missed. The critical switching fields (which reduce the barriers to zero) range from 8--16 MV/cm, higher than the values (5--8 MV/cm) estimated from finite-temperature MD simulations. This outcome aligns with the expected trend, as DFT calculations assuming homogeneous switching often overestimate the critical fields~\cite{Beckman09p144124}. (\textbf{b}) Polarization variation along SA and SI-1 pathways from the same starting configuration (top), corresponding to two distinct $P-\mathcal{E}$ hysteresis loops (bottom left). Strain ($\eta_3$) as a function of an electric field applied along the $z$ axis ($\mathcal{E}_3$). The sign of $d_{33}$ (the slope of $\eta_3$-$\mathcal{E}_3$ curve) depends on the arbitrary choice of the sign of $\mathcal{E}_3$. Reproduced with permission from ref.~\cite{Ma23p256801}.}
\label{fig_mobility}
\end{figure}

The alternating \Op~and \Onp~layers result in multiple polarization switching pathways, even at the unit cell level~\cite{Qi22parXiv,Choe21p8,Wei22p154101,Ma23p256801,Dou24p092901}. These pathways can be classified as shift-inside (SI) and shift-across (SA). Specifically, the SI-1 pathway involves only the displacement of \Op~atoms, with the transition state adopting the $T$ phase ($P4_2/nmc$). In the SI-2 pathway, both \Op~and \Onp~ions move against the electric field ($\mathcal{E}$), resulting in a coordinated transition of \Onp$\rightarrow$\Op~and \Op$\rightarrow$\Onp. In contrast, \Op~ions move across through Hf planes in the SA pathway. The energy barriers of SI-1, SI-2 and SA, calculated by DFT-based VCNEB and with the $X_2^-$ mode conserved, are 0.39, 0.22 and 0.79 eV/unitcell, respectively (Figure~\ref{fig_mobility}\textbf{a})~\cite{Ma23p256801}. We make a brief comment on the subtle differences between the SI-2 pathways as identified with VCNEB and NEB. The former technique, which allows for lattice relaxation, predicts a $Pbcn$-like intermediate state~\cite{Ma23p256801}, whereas the latter identifies the $T$ phase as the intermediate and yields a higher barrier of 0.34 eV/unitcell. Therefore, the barrier calculated using VCNEB represents a lower limit for homogeneous polarization switching, whereas the estimate obtained with NEB approximates the upper limit. The strain-dependent stability of the $Pbcn$ phase was also discussed based on symmetry mode analysis~\cite{Zhou22peadd5953}.

When comparing the SI and SA pathways, it is observed that the same initial structure can undergo polarization switching driven by electric fields in opposite directions~\cite{Choe21p8,Ma23p256801}. To conform with classical electrodynamics, the same structure would exhibit polarization along opposite direction in the SI and SA pathways. This counter-intuitive consequence is actually a manifestation of the geometric-quantum-phase nature of electric polarization ($P$), a multi-valued \textit{lattice} property without a specific direction. The SI and SA pathways correspond to two branches of the polarization lattice, each associated with a well-defined polarization change $\Delta P$ without ambiguity (Figure~\ref{fig_mobility}\textbf{b}). Consequently, \citeauthor{Ma23p256801} proposed that HfO$_2$ exhibits dual-valued remnant polarization ($P_s^{\rm SI}=-0.5$ C/m$^2$ and $P_s^{\rm SA}=0.7$ C/m$^2$), each with distinct $P-\mathcal{E}$ hysteresis loop\cite{Ma23p256801}. \citeauthor{Wu23p226802} argued that HfO$_2$ should exhibit an intrinsic polarization value of 0.7 C/m$^2$ based on the domain wall motion achieved through an SA-like mechanism, although the barrier is rather high at 0.5 eV~\cite{Wu23p226802}.

An interesting question naturally arises given the dual-valued remnant polarization: do the two switching pathways correspond to two distinct values of piezoelectric strain coefficient $d_{33}$? \citeauthor{Liu20p197601} first predicted that $Pca2_1$ HfO$_2$ exhibits a negative value of $d_{33}$ such that the lattice along the polar axis shrinks when an electric field is applied along the direction of the polarization ($\mathcal{E}_3$). Moreover, both transverse and longitudinal piezoelectric coefficients are negative, leading to an unusual ``electric auxetic effect," where the lattice will shrink (expand) along all three dimensions when the field is applied along (against) the direction of $P$. Experimentally, both positive and negative coefficients have been observed in ferroelectric hafnia-based thin films~\cite{Dutta21p7301,Starschich14p202903,Liu23p172083}. Some studies proposed that the double-path polarization switching in hafnia is responsible for the varying piezoelectric responses observed~\cite{Qi22parXiv}. Based on deep-learning-assisted finite-field molecular dynamics simulations, \citeauthor{Ma23p256801} reported that an electric field driving \Op~ions away from their nearest Hf atomic layer will result in lattice expansion ($\eta_3>0$), and vice versa (Figure~\ref{fig_mobility}\textbf{b}). Thus, for a given crystal orientation and $\mathcal{E}_3$, the induced strain change is unique; the absolute value of $d_{33}$ calculated using $\lvert \partial \eta_3$/$\partial \mathcal{E}_3 \rvert$ is single-valued, with the sign of $d_{33}$ entirely dependent on the sign of $\mathcal{E}_3$ (the choice of the positive direction of the field is arbitrary)~\cite{Ma23p256801}. Another way to understand this is to recognize that the two branches of the polarization lattices corresponding to the SI and SA pathways have the same slope, supporting a singe-valued $|d_{33}|$. Since polarization itself does not have a well defined sign due to its geometric-quantum-phase nature, we cannot state that an electric field (a vector quantity) is ``aligned along" the polarization (a multi-valued lattice property). However, the relative orientation between a field and crystal orientation can, in principle, be determined, and the resulting strain change is thus well-defined and unambiguous. Interestingly, the combined use of band-excitation piezoresponse force microscopy and molecular dynamic simulations demonstrated that the $d_{33}$ of ZrO$_2$-based antiferroelectrics exhibits intriguing behavior where its sign changes with an increasing electric field, arsing from the field-driven $T\rightarrow Pca2_1$ phase transition~\cite{Richard22p054066,Lomenzo23p2303636}.

The long-distance travel of oxygen ions through successive SI and SA transitions is similar to adiabatic Thouless pumping, achievable by applying a constant bias. Large-scale finite-field MD simulations have confirmed that the presence of a unidirectional electric field can drive successive ferroelectric switching, supporting a continuous flow of oxygen ions and high oxygen ion mobility at moderate temperatures, even in the absence of oxygen vacancies~\cite{Ma23p256801}. Investigating the potential interaction between ferroelectric switching and ion transport at high fields might be crucial for the development of novel hafnia-based nanoelectronics and electric-field-assisted ultrafast ion conductors.

\subsection{Domain wall motion}

Domain walls (DWs) are interfaces separating domains with different polarization directions. The dynamics of DWs in perovskite ferroelectrics have been extensively studied, and it is now widely acknowledged that the mobility of DWs dictates the switching speed and coercive field. \citeauthor{Kiguchi18p11UF16} identified several types of 90\degree~and 180\degree~DWs in epitaxially grown Y-doped HfO$_2$ (YHO) films using scanning transmission electron microscopy~\cite{Kiguchi18p11UF16}. Understanding the atomic-scale structural properties and dynamic response of DWs in ferroelectric HfO$_2$ is essential for addressing several critical challenges, particularly the giant coercive field that is orders of magnitude higher than their ferroelectric counterparts.

\begin{figure}[t]
\includegraphics[width=1.0 \textwidth]{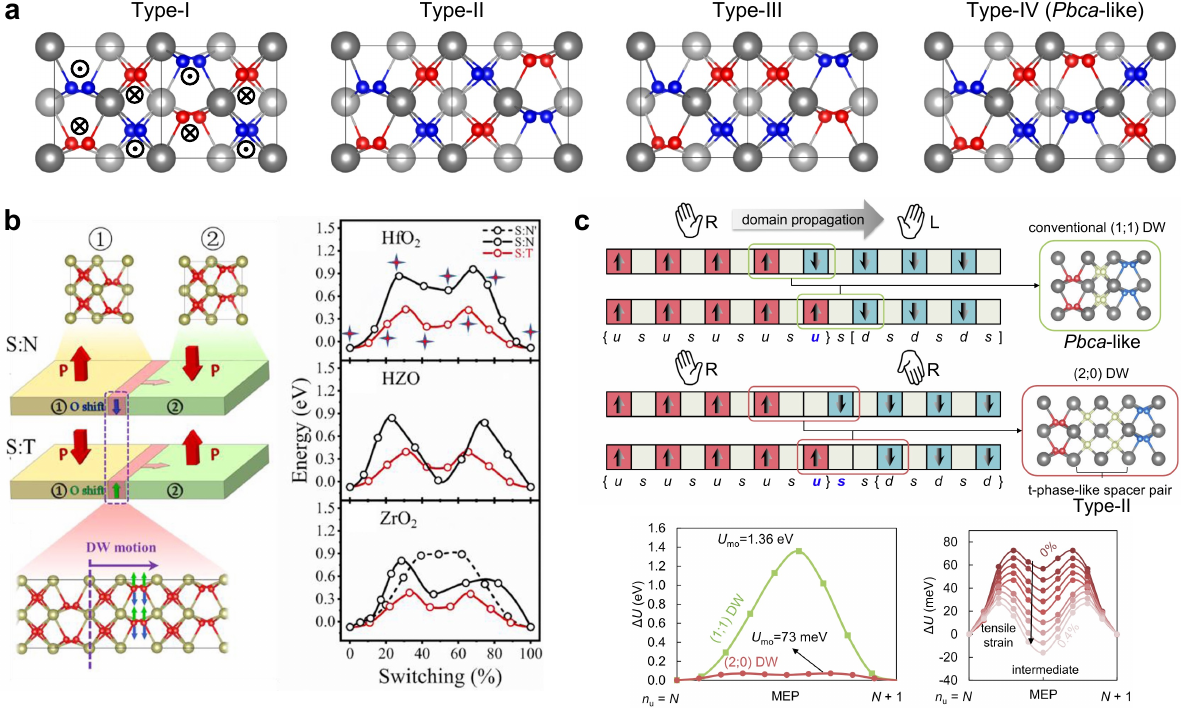}
\caption{Domain-wall motions in $Pca2_1$ HfO$_2$. (\textbf{a}) Four types of 180\degree~DWs classified by whether the sign of the $X_2^-$ mode is conserved (type-I and type-II) or reversed (type-III and type-IV) across the boundary. The type-IV wall is similar to the antipolar $Pbca$ phase. The oxygen atoms are colored according to the direction of their $x$-displacements. Walls with oxygen atoms of the same color indicate a reversal of the $X_2^-$ mode. (\textbf{b}) Sketch of the motion for a $Pbca$-type 180\degree~DW along the non-crossing (S:N, SI-1-like in Figure~\ref{fig_mobility}) and crossing (S:T, SA-like in Figure~\ref{fig_mobility}) switching pathways (left) and energy variation during DW motion in HfO$_2$, HZO, and ZrO$_2$, respectively (right). (\textbf{c}) Schematics for the motions of a (1;1) DW, same as the $Pbca$-type wall, and a (2;0) DW, similar to the type-II wall in (\textbf{a}). The barrier for the motion of a (2;0) wall is much lower than that of the conventional $Pbca$-type wall with the $X_2^-$ mode reversed (bottom left), and can be further reduced by applying tensile strain (bottom right). Panels \textbf{a}, \textbf{b}, and \textbf{c} are reproduced with permissions from refs.~\cite{Qi21parXiv}, \cite{Wu23p226802}, and~\cite{Choe21p8}, respectively.}
\label{fig_dw}
\end{figure}

The unit cell of $Pca2_1$ HfO$_2$ exhibits conditional chirality: the application of Euclidean transformations within the unit cell results in eight different representations~\cite{Choe21p8}. This leads to a rich spectrum of 180$^\circ$ DWs. One of the most studied DW type is the $Pbca$-type wall, which is thermodynamically the most stable ~\cite{Lee20p1343,Ding20p556,Choe21p8,Wu23p226802} and resembles the anti-polar $Pbca$ phase at the boundary that separates oppositely polarized domains. \citeauthor{Lee20p1343} simulated the motion of a $Pbca$-type wall with DFT-based NEB, assuming the wall moves via the SI-1-like mechanism during which the \Op~atoms move between Hf atomic players, referred to as ``non-crossing" mechanism in later studies. However, the calculated barrier is quite high, at $\approx 1.3$~eV. \citeauthor{Silva23p101064} identified a ``crossing" pathway, during which the $Pbca$-type wall propagates via an SA-like mechanism with \Op~atoms moving across the Hf atomic planes. The crossing pathway has a lower barrier than the non-crossing one, and La doping can further reduce the barrier to 0.4~eV~\cite{Silva23p101064}. Recently, \citeauthor{Wu23p226802} reported similar results for $Pbca$-type walls (0.51 eV for crossing and 1.04 eV for non-crossing) (Figure~\ref{fig_dw}\textbf{b}) and then proposed that ferroelectric HfO$_2$ should exhibit a higher polarization value of 0.7 C/m$^2$ than the conventional value of 0.5 C/m$^2$ deduced from the SI-like switching pathway. However, we note that the barrier of the crossing pathway in the motion of the $Pbca$-type wall is substantially higher than the SI-2 pathway for homogeneous domain switching (0.22 eV), casting doubts on their relevance to the switching behavior of HfO$_2$. 

The reason the non-crossing pathway for the motion of a $Pbca$-type wall must overcome such a large barrier is due to the requirement to flip the sign of the $X_2^-$ mode. \citeauthor{Qi21parXiv} investigated the switching barriers for four different types of 180\degree~DWs in HfO$_2$, distinguished by whether the $X_2^-$ mode conserves or reverses its sign across the wall (Figure~\ref{fig_dw}\textbf{a})~\cite{Qi21parXiv}. It was pointed out that the $Pbca$-type wall has the sign of the $X_2^-$ mode reversed across the wall. The MEPs identified with DFT-based NEB calculations confirmed that the walls conserving the $X_2^-$ sign have much lower switching barriers than those reversing it. This occurs because the $X_2^-$ mode features anti-parallel $x$-displacements of oxygen atoms, orthogonal to the direction of the polarization. This mode, not involving a net change in the local dipole moment, cannot be driven by an electric field applied orthogonally to the polarization direction. \citeauthor{Choe21p8} also found that the $X_2^-$-sign-conserved walls are much more mobile than $Pbca$-type walls (Figure~\ref{fig_dw}\textbf{c}). We note that these results also help explain the lower barrier associated with the crossing pathway for the motion of a $Pbca$-type wall: the oxygen atoms preserve their respective signs of $x$-displacements while moving across the Hf planes. Recently, \citeauthor{Xu24p012902} looked into the structural and electronic properteis of oxygen-vacacany-stablized head-to-head and tail-to-tail charged DWs of ferroelectric HfO$_2$~\cite{Xu24p012902}. The non-180\degree~DWs in HfO$_2$ are less explored computationally. \citeauthor{Ding20p556} computed the lateral motion barriers for several types of 90\degree~DWs, although these walls reside in the (001) planes instead of the (011) planes commonly found in perovskite ferroelectrics like PbTiO$_3$.

These DFT-based computational studies have revealed extraordinary mechanistic and energetic complexities in the motion of DWs in ferroelectric HfO$_2$. Although not directly related to polarization, the $X_2^-$ mode plays a significant role in both polymorphism and switching kinetics. It is well established that directly estimating the coercive field from the switching barrier typically overestimates the value by several orders of magnitude~\cite{Beckman09p144124}. Gaining a deeper understanding of the thermally activated nucleation and growth~\cite{Shin07p881, Liu16p360} processes at the wall is necessary for making accurate quantitative predictions of coercive fields in the future.

\section{MODELING BEYOND DFT}

In previous discussions, we have mainly focused on results obtained with zero-Kelvin DFT calculations. It has become evident that the properties of HfO$_2$-based ferroelectric devices are strongly influenced by various kinetic and dynamic processes that span multiple time and spatial scales~\cite{Park18p1800522,Park18p716}. Particularly during phase transitions, kinetic effects such as those occurring in annealing and cooling processes are critical for forming and stabilizing the metastable ferroelectric $Pca2_1$ phase. The dynamical processes involved in domain and DW switching govern the switching speed and coercive field, thereby affecting write speed and power consumption. ~\cite{Liu16p360,Li19p126502}. A deeper understanding of the phase transition kinetics and polarization switching dynamics in HfO$_2$ and its doped variants could hold the key to improving the viability of hafnia-based devices. The high computational cost of DFT limits the exploration of finite-temperature dynamical properties. We suggest molecular dynamics (MD) simulations and phase-field modeling are valuable approaches to overcome these limitations.

\subsection{Deep potential molecular dynamics}

In classical MD simulations, atoms are treated as structureless classical particles that interact through a classical force field and obey Newton’s laws of motion. This approach efficiently simulates the time-dependent behavior of an ensemble of atoms. The accuracy of the force fields, which describe the interatomic interactions, is critical to the success of MD simulations. Historically, HfO$_2$ has been extensively studied for decades as a gate dielectric material and for its potential applications in resistive switching memory, long before its ferroelectric properties in thin films were discovered. During this period, several classical force fields of hafnia were developed~\cite{Shan10p125328,Wang12p224110,Trinastic13p154506}. These conventional force fields employ a set of analytical energy functions to model interatomic interactions, with parameters fitted to align the simulation results with known reference data. However, these force fields were developed without knowledge of the then-unknown ferroelectric $Pca2_1$ phase, which calls into question their effectiveness in studying the ferroelectric properties of $Pca2_1$ HfO$_2$. Machine learning (ML) methods, known for their sophisticated handling of complex, high-dimensional data, provide promising avenues for developing force fields that are both transferable and scalable, with a precision comparable to DFT. Here, we focus on a specific machine-learning force field (MLFF), namely Deep Potential (DP), which employs deep learning techniques to train deep neural networks using first-principles data.

\begin{figure}[t]
\includegraphics[width=0.7 \textwidth]{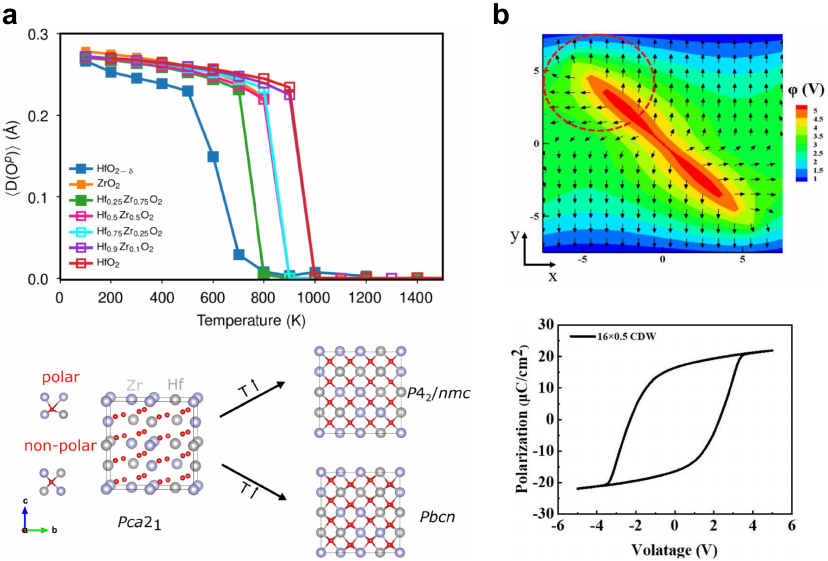}
\caption{Modeling HfO$_2$ beyond DFT. (\textbf{a}) Two distinct phase transitions starting from the $Pca2_1$ phase revealed by large-scale MD simulations using a deep potential of Hf$_x$Zr$_{1-x}$O$_{1-\delta}$. The solid and empty markers denote $Pca2_1$ $\rightarrow$ $P4_2/nmc$ and $Pca2_1$ $\rightarrow$ $Pbcn$, respectively. (\textbf{b}) Phase-field modeling of $P$-$V$ hysteresis for HZO thin films containing 90\degree~tail-to-tail DWs. Panels \textbf{a} and \textbf{b} are reproduced with permissions from refs.~\cite{Wu23p144102}, and~\cite{Zhou22p117920}, respectively.} 
\label{fig_dp}
\end{figure}

\citeauthor{Wu21p024108} developed a DP model of HfO$_2$ using a concurrent learning procedure, which enabled accurate predictions of various structural properties such as elastic constants and equations of states for different hafnia polymorphs, phonon dispersion relationships, and phase transition barriers, achieving a level of accuracy comparable to DFT calculations~\cite{Wu21p024108}. They further developed a unified model potential applicable to Hf$_x$Zr$_{1-x}$O$_2$ solid solutions across a range of $x$ values, employing the strategy of ``modular development of deep potential”~\cite{Wu23p144102}. Using this unified DP model, a series of constant-temperature constant-pressure ($NPT$) MD simulations were performed to study composition-dependent, temperature-driven phase transitions. Intriguingly, MD simulations uncovered two distinct phase transitions: $Pca2_1\rightarrow P4_2/nmc$ in Hf$_x$Zr$_{1-x}$O$_2$ with $x$ = 0.0, 0.25, and $Pca2_1 \rightarrow Pbcn$ in Hf$_x$Zr$_{1-x}$O$_2$ with $x$ = 0.5, 0.75, 0.9, 1.0. The $Pbcn$ phase resembles an antiferroelectric, characterized by neighboring antiparallel \Op~atoms. While the transition from $Pca2_1$ to $P42/nmc$ is commonly associated with the temperature-driven polar-to-nonpolar transition in ferroelectric hafnia and zirconia thin films (Figure~\ref{fig_dp}\textbf{a}), the $Pbcn$ phase has been hypothesized as an intermediate orthorhombic. Recently, based on MD simulations with a DP model of HfO$_2$, \citeauthor{Ma23p256801} discovered a bias-induced ultrahigh oxygen ion mobility at moderate temperatures in ferroelectric HfO$_2$~\cite{Ma23p256801}, highlighting the ability of MD simulations to offer detailed atomistic insights without any \textit{a priori} assumptions.\\

\subsection{Phase-field simulations}

Phase-field modeling is the most popular and powerful method for modeling ferroelectrics at the mesoscopic length scale. This approach, rooted in Ginzburg-Landau theory, represents the free energy continuously via the order parameter. These thermodynamic functions, typically polynomial expressions of independent variables, could be extended to both homogeneous and inhomogeneous phases. The spatial distribution of the order parameter governs the inhomogeneous properties at the interface, and the free energy is calculated by summing all energy contributions associated with the order parameter. The evolution over time of the order parameter is described by the time-dependent Ginzburg-Landau equations~\cite{Chen08p1835}.

\citeauthor{Noh19p486} performed phase-field simulations on HZO, utilizing a Landau free energy expansion up to the 6th order terms with Landau coefficients calibrated to experimental data. They observed that the DWs could spontaneously move even without an applied electric field, leading to spontaneous polarization excitation and relaxation processes critical for accumulative $P$-switching~\cite{Saha19p202903}. \citeauthor{Zhou22p117920} explored the impact of 90\degree~tail-to-tail DWs on the wake-up effects in polycrystalline HZO films (Figure~\ref{fig_dp}\textbf{b})~\cite{Zhou22p117920}, using Landau coefficients derived from $P$-$\mathcal{E}$ hysteresis data~\cite{Kao18p4652,Chang20pSGGA07}. \citeauthor{Sugathan22p14997} used a genetic algorithm to optimize the effective Landau coefficients describing the free energy of HZO, reproducing experimental $P$-$\mathcal{E}$ curves for polycrystalline thin films~\cite{Sugathan22p14997}. Findings from phase-field simulations suggested that grain morphology and crystalline texture can be unitized to optimize the ferroelectric properties of HZO thin films. 

Despite these efforts, two significant challenges persist in ensuring reliable phase-field simulations of hanifa-based thin films. Firstly, given the presence of multiple polymorphs, which result from combinations of tetragonal, polar, and antipolar modes (see the section `Symmetry mode analysis of $Pca2_1$'), the order parameters adopted in the Landau free energy fundamentally restrict the types of polymorphs that can be represented in phase-filed simulations. Recent progress includes the introduction of an order parameter to differentiate between the crystal and the amorphous phases~\cite{Liu24p44}. However, additional order parameters are needed to describe competing nonpolar phases such as $M$ and $T$. Another challenge in applying phase-field simulations to polycrystalline multiphase hafnia-based ferroelectric thin films lies in establishing a uniform set of parameters, similar to those used for traditional perovskite ferroelectric materials.

\section{SUMMARY AND OUTLOOK}

In this review, we have highlighted the computational efforts made in elucidating the ferroelectric mechanisms of HfO$_2$, a prominent candidate for realizing the long-pursued goal of integrating ferroelectric functionality into integrated circuits to make ferroelectric memory commercially viable. We have examined various polar phases proposed to explain the ferroelectricity in hafnia-based thin films, assessing their stability, polarization characteristics, and experimental supporting evidences. The $Pca2_1$ phase appears to be the one most compatible with both DFT-based analyses and experimental findings, though ongoing debates are anticipated due to HfO$_2$'s diverse polymorphs and their extrinsic factor-dependent stability.

We have seen extensive DFT-based investigations, exploring stabilizing factors essential for the formation of the metastable $Pca2_1$ phase. Although the surface energy model provides a plausible explanation for the inverse size effect, computing reliable surface energies for HfO$_2$ polymorphs remains a challenge and consequently a worthy direction for future research. While DFT modeling of epitaxial strain effect has demonstrated remarkable success in the studies of epitaxial thin films of perovksite ferroelectrics, its significance in polycrystalline or epitaxial hafnia thin films, which often contain stress-relieving defects or undergo domain-matching epitaxy, is still up for debate. 

The role of charged oxygen vacancies in structural polymorphism kinetics revealed by DFT calculations underscores the unique aspect of ferroelectric hafnia: the phase transition kinetics are at least as important as polarization switching kinetics given their similar energy barriers. These vacancies, like a double-edged sword, facilitate both nonpolar-to-polar and polar-to-nonpolar phase transitions. Recent experiments demonstrated that the acceptor-donor co-doping, which regulates the level of oxygen vacancy, is a promising approach to improve the ferroelectric properties of HfO$_2$~\cite{Zhou24p2893}.

There are many open questions, especially regarding the polarization switching mechanisms involving domains and domain walls and their quantitative connections to macroscopic coercive fields. Traditional zero-Kelvin DFT-based nudged elastic band techniques, which require manual pathway construction, might introduce bias when determining the ferroelectric switching mechanisms. We suggest large-scale MD simulations at finite temperatures and in the presence of electric fields, assisted by machine-learning force fields, could be a powerful tool to accurately determine the mechanisms at the atomic level without \textit{a priori} assumption. Those computationally efficient force fields applicable to dopants and vacancies, if developed, could also facilitate a statistical analysis of the doping effect on the stability of hafnia polymorphs. 

A significant gap exists between the prevalent interest in polycrystalline hafnia thin films used in device applications and the computational focus on single-crystal forms. Modeling polycrystalline films, which feature varied phases of HfO$_2$ and structural defects, is exceptionally challenging. A multiscale approach that combines DFT, MD, and phase-field modeling might not only be beneficial but necessary for achieving a correct mechanistic understanding of ferroelectric behavior of HfO$_2$ under device-relevant conditions. Despite ongoing debates over the ferroelectric mechanisms in HfO$_2$, one thing becomes evident: we are entering a new era of ferroelectric physics, thanks to this fluorite-structured, not-so-simple binary oxides. For computational physicists and materials scientists focused on ferroelectricity, it is exhilarating to be part of what remains a golden era, over a century since the discovery of the phenomenon.
\\

{\bf{Data Availability}} Data sharing is not applicable to this article as no datasets were generated or analyzed during the current study.

{\bf{Acknowledgments}} T.Z., L.M., S.D., and S.L. acknowledge the supports from National Key R\&D Program of China (2021YFA1202100), National Natural Science Foundation of China (12361141821, 12074319), and Westlake Education Foundation. The computational resource is provided by Westlake HPC Center.

{\bf{Competing Interests}} The authors declare no competing financial or non-financial interests.

{\bf{Author Contributions}} S.L. led the collaboration. T.Z. developed the manuscript outline. T.Z. and L.M. wrote a major part of the manuscript, and S.L. revised the manuscript entirely. T.Z. and L.M. contributed equally to this work. All authors reviewed and edited the manuscript.

\clearpage
\bibliography{SL}

\clearpage

\clearpage
\begin{table}[t]
\caption{DFT lattice parameters, relative energies, and polarization values of different HfO$_2$ phases. The lattice constants ($a$, $b$, $c$), angles ($\alpha$, $\beta$, $\gamma$), energy ($E$), and polarization ($P$) are in units of \AA, degrees, meV/f.u., and $\mu$C/cm$^2$, respectively. The $R3m$ and $R3$ phases can be modeled using either a 12-atom rhombohedral (pseudocubic) unit cell or a 36-atom hexagonal unit cell. The Perdew-Burke-Ernzerhof (PBE) exchange correlation functional~\cite{Perdew96p3865} and a plane-wave cutoff energy of 600 eV are employed. The Brillouin zones of the 12-atom pseudocubic cell, 24-atom $Pbca$ or $Pc$ cell, and 36-atom hexagonal cell are sampled by $\Gamma$-centered ($4\times4\times4$), ($2\times4\times4$), and ($3\times3\times3$) Monkhorst-Pack~\cite{Monkhorst76p5188} $k$-point meshes, respectively.}
\begin{tabular}{p{1.5 in}p{0.5 in}<{\centering}p{0.5 in}<{\centering}p{0.5 in}<{\centering}p{0.5 in}<{\centering}p{0.5 in}<{\centering}p{0.5 in}<{\centering}p{0.5 in}<{\centering}p{0.5 in}<{\centering}}
\hline
\hline
Phase  &  $a$  &  $b$  &  $c$  &  $\alpha$  &  $\beta$  &  $\gamma$  &  $E$  &  $P$  \\
\hline
$P2_1/c$ ($M$)            &  5.14  &  5.19  &  5.32  &  90.0  &  99.7  &  90.0  &     0  &     0  \\
$Pbca$                    & 10.18  &  5.17  &  5.32  &  90.0  &  90.0  &  90.0  &  27.4  &     0  \\
$Pc$                      & 10.14  &  5.24  &  5.17  &  90.0  &  94.8  &  90.0  &  65.4  &  27.2  \\
$Pbca$ ($AO$)             & 10.05  &  5.08  &  5.25  &  90.0  &  90.0  &  90.0  &  72.9  &     0  \\
$Pca2_1$                  &  5.05  &  5.08  &  5.27  &  90.0  &  90.0  &  90.0  &  84.3  &  50.3  \\
$Pmn2_1$                  &  5.12  &  5.12  &  5.18  &  90.0  &  90.0  &  84.3  & 142.9  &  56.2  \\
$P4_2/nmc$ ($T$)          &  5.08  &  5.08  &  5.23  &  90.0  &  90.0  &  90.0  & 166.2  &     0  \\
$P\bar{4}3m$ (cubic)      &  5.10  &  5.10  &  5.10  &  90.0  &  90.0  &  90.0  & 204.2  &     0  \\
$P\bar{4}3m$ (hexagonal)  &  7.21  &  7.21  &  8.83  &  90.0  &  90.0  & 120.0  & 204.2  &     0  \\
$R3m$ (rhombohedral)      &  5.10  &  5.10  &  5.10  &  90.0  &  90.0  &  90.0  & 204.2  &   0.1  \\
$R3m$ (hexagonal)         &  7.21  &  7.21  &  8.83  &  90.0  &  90.0  & 120.0  & 204.2  &   0.1  \\
$R3$ (rhombohedral)       &  5.14  &  5.14  &  5.14  &  88.7  &  88.7  &  88.7  & 215.9  &  41.4  \\
$R3$ (hexagonal)          &  7.18  &  7.18  &  9.09  &  90.0  &  90.0  & 120.0  & 215.9  &  41.4  \\
$Fm\bar{3}m$              &  5.07  &  5.07  &  5.07  &  90.0  &  90.0  &  90.0  & 269.7  &     0  \\
$R3m$ (5\%-strained)      &  6.85  &  6.85  &  9.86  &  90.0  &  90.0  & 120.0  & 297.0  &  24.6  \\
\hline
\hline
\end{tabular}
\label{tab_phase}
\end{table}

\clearpage

\clearpage

\clearpage

\clearpage
\begin{table}[t]
\caption{Calculated surface energies of different HfO$_2$ phases for various surface planes. The values, in unit of J/m$^2$, outside and inside the parentheses are extracted from refs.~\cite{Batra16p172902} and~\cite{Wu20p252904}, respectively.}
\begin{tabular}{p{0.6 in}<{\centering}p{1.0 in}<{\centering}p{1.0 in}<{\centering}p{1.0 in}<{\centering}p{0.6 in}<{\centering}}
\hline
\hline
Surface  &  $P2_1/c$  &  $P4_2/nmc$  &  $Pca2_1$  &  $Pmn2_1$  \\
\hline
(001)  &  1.51 (1.37)  &  1.21 (1.03)  &  1.41 (1.24)  &  0.79  \\
(010)  &  1.88 (1.79)  &  1.55 (1.60)  &  2.68 (2.61)  &  2.34  \\
(100)  &  1.67 (1.61)  &  1.55 (1.60)  &  1.83 (1.80)  &  2.34  \\
(110)  &  1.38 (1.34)  &  1.08 (1.05)  &  1.98 (1.94)  &  2.09  \\
(1$\bar{1}$0)  &  1.38  &  1.08  &  1.98  &  1.69  \\
(101)  &  1.57 (1.53)  &  1.54 (1.54)  &  1.61 (1.59)  &  1.67  \\
(011)  &  (1.45)  &  (1.54)  &  (2.10)  &  ---  \\
(111)  &  1.25 (1.21)  &  1.12 (1.13)  &  (1.40)  &  ---  \\
\hline
\hline
\end{tabular}
\label{tab_surface}
\end{table}

\clearpage

\clearpage

\clearpage

\clearpage

\clearpage

\clearpage

\clearpage

\end{document}